\documentclass[a4paper,11pt, fleqn]{article}

\usepackage{enumitem}

\usepackage[a4paper, total={6.5in, 10in}]{geometry}
\usepackage{graphicx}
\usepackage{amsmath}
\usepackage{amssymb}
\usepackage{upgreek}
\usepackage{multicol}
\usepackage{color}
\usepackage{changepage}
\usepackage{booktabs}
\usepackage{multirow}
\usepackage{siunitx}
\usepackage{rotating}
\usepackage{float}
\usepackage{Tabbing}
\usepackage{dsfont}
\usepackage{amsthm}
\usepackage{amstext}
\usepackage{hyperref}
\usepackage{cleveref}
\usepackage[normalem]{ulem}
\usepackage{array}
\usepackage{caption}
\usepackage{graphicx}
\graphicspath{ {./images/} }
\usepackage[english]{babel}
\usepackage[utf8]{inputenc}
\usepackage[round]{natbib}
\usepackage[title]{appendix}
\usepackage{authblk}

\crefname{lemma}{Lemma}{Lemmas}
\crefname{corollary}{Corollary}{Corollaries}
\crefname{theorem}{Theorem}{Theorems}
\crefname{definition}{Definition}{Definitions}
\crefname{example}{Example}{Examples}
\crefname{remark}{Remark}{Remarks}
\crefname{assumption}{Assumption}{Assumptions}

\newtheorem{theorem}{Theorem}
\newtheorem{assumption}{Assumption}

\newtheorem{lemma}{Lemma}
\newtheorem{example}{Example}

\renewcommand{\thesection}{\arabic{section}}
\renewcommand{\theequation}{\arabic{section}.\arabic{equation}}

\renewcommand{\theassumption}{\arabic{section}.\arabic{assumption}}

\renewcommand{\thelemma}{\arabic{section}.\arabic{lemma}}

\newcommand{\abs}[1]{\left\lvert#1\right\rvert}
\newcommand{\norm}[1]{\left\lVert#1\right\rVert}
\newcommand{\E}{\mathbb{E}}
\renewcommand{\P}{\mathbb{P}}

\newcommand{\setEP}[1]{\mathcal{E}_{#1}} % Notation for empirical process sets
\newcommand{\setCC}{\mathcal{CC}_T} % Notation for covariance closeness sets
\newcommand{\setEPl}[1]{\setEP{T}^l}

\newcommand{\bzero}{\boldsymbol{0}}
\newcommand{\bA}{\boldsymbol{A}}
\newcommand{\bepsilon}{\boldsymbol{\epsilon}}
\newcommand{\bzeta}{\boldsymbol{\zeta}}
\newcommand{\by}{\boldsymbol{y}}
\newcommand{\bx}{\boldsymbol{x}}
\newcommand{\bq}{\boldsymbol{q}}
\newcommand{\bu}{\boldsymbol{u}}
\newcommand{\bv}{\boldsymbol{v}}

\newcommand{\bb}{\boldsymbol{b}}
\newcommand{\bB}{\boldsymbol{B}}
\newcommand{\be}{\boldsymbol{e}}
\newcommand{\bz}{\boldsymbol{z}}

\newcommand{\br}{\boldsymbol{r}}
\newcommand{\bbf}{\boldsymbol{f}}

\newcommand{\bX}{\boldsymbol{X}}
\newcommand{\bW}{\boldsymbol{W}}
\newcommand{\bR}{\boldsymbol{R}}
\newcommand{\bw}{\boldsymbol{w}}

\newcommand{\bI}{\boldsymbol{I}}
\newcommand{\cH}{\mathcal{H}}

\newcommand{\bDelta}{\boldsymbol{\Delta}}
\newcommand{\bbeta}{\boldsymbol{\beta}}
\newcommand{\bmu}{\boldsymbol{\mu}}
\newcommand{\bxi}{\boldsymbol{\xi}}
\newcommand{\bXi}{\boldsymbol{\Xi}}

\newcommand{\bgamma}{\boldsymbol{\gamma}}
\newcommand{\bPhi}{\boldsymbol{\Phi}}
\newcommand{\bpsi}{\boldsymbol{\psi}}
\newcommand{\bM}{\boldsymbol{M}_{(\boldsymbol{X}_\cH)}}

\newcommand{\bSigma}{\boldsymbol{\Sigma}}
\newcommand{\bOmega}{\boldsymbol{\Omega}}
\newcommand{\bTheta}{\boldsymbol{\Theta}}
\newcommand{\bUpsilon}{\boldsymbol{\Upsilon}}
\newcommand{\bGamma}{\boldsymbol{\Gamma}}
\newcommand{\Weight}{\boldsymbol{W}}
\newcommand{\diag}{\text{diag}}
\newcommand{\bLambda}{\boldsymbol{\Lambda}}

\newcommand{\bH}{\boldsymbol{H}}

\newcommand{\bC}{\boldsymbol{C}}
\newcommand{\bF}{\boldsymbol{F}}

\newcommand{\probseqCC}{\upsilon^{\mathcal{CC}}_{N,T}}
\newcommand{\probseqE}{\upsilon^{\mathcal{E}}_{N,T}}

\DeclareMathOperator*{\argmin}{arg\,min}

\title{Local Projection Inference in High Dimensions}
\author{Robert Adamek$^\dagger$, Stephan Smeekes$^\ddagger$, Ines Wilms$^\ddagger$\\
$^\dagger$Department of Economics and Business Economics, Aarhus University,\\ Fuglesangs All\'e 4, 8210 Aarhus V, Denmark\\
Email: \texttt{r.adamek@econ.au.dk}\\
{$^\ddagger$Department of Quantitative Economics, Maastricht University,\\ Tongersestraat 53, 6211LM, The Netherlands}\\
Email: \texttt{\{s.smeekes,i.wilms\}@maastrichtuniversity.nl}
}
\date{\today}

\usepackage[nodisplayskipstretch]{setspace}
\begin{document}
\onehalfspacing
\maketitle
\begin{abstract}
In this paper, we estimate impulse responses by local projections in high-dimensional settings. We use the desparsified (de-biased) lasso to estimate the high-dimensional local projections, while leaving the impulse response parameter of interest unpenalized. We establish the uniform asymptotic normality of the proposed estimator under general conditions. Finally, we demonstrate small sample performance through a simulation study and consider two canonical  applications in macroeconomic research on monetary policy and government spending.
\bigskip

\end{abstract}

% main body

\section{Introduction}\label{sec:Introduction}
In this paper, we develop a simple approach for conducting valid inference on impulse responses in high-dimensional settings. We do so by pairing the estimation framework of local projections (LPs) with the inference framework of the desparsified (or debiased) lasso. Since their introduction by \citet{jorda2005estimation}, LPs have been widely used in macroeconomic research for studying the dynamic propagation of shocks through impulse response analysis, see e.g. \citet{angrist2018semiparametric}, \citet{ramey2018government}, \citet{ramey2016macroeconomic} and \citet{stock2018identification}. 
As such, they have become
an increasingly used alternative to structural Vector AutoRegressions (SVAR) pioneered by \citet{sims1980macroeconomics}, 
see for instance \citet{ramey2016macroeconomic} or \citet{kilian2017structural} for an overview on this work.

While the methods have different finite-sample properties, SVARs and LPs are equivalent in population, as established by \cite{plagborg2021local} who show that the underlying impulse response estimands are the same for both. This implies that the well-documented necessity for SVARs to add assumptions in order to achieve structural identification of the impulse responses, is equally true for LPs. Indeed, \cite{plagborg2021local} show that the identification strategies typically used for SVARs have an equivalent implementation for LPs, and vice versa. While identification issues therefore do not affect the choice between SVAR or LP, finite-sample considerations about estimation and inference problem do. LP impulse responses are obtained by estimating only univariate linear regressions, and performing standard inference on (typically) a single parameter of interest across these univariate regressions.  In contrast, SVARs require estimating the whole system of equations, and transforming these into the Vector Moving Average (VMA) representation from which the impulse responses can be derived. While the system estimation of the SVAR can still be done linearly, the inversion to the VMA representation is a nonlinear operation that renders the impulse response coefficients a complex nonlinear functions of \emph{all} VAR parameters. This makes inference considerably more cumbersome, even in
low-dimensional settings, where the complications and inaccuracies of applying the Delta method in finite samples have led to bootstrap inference becoming the norm \citep[see e.g. Chapter 12 of][and the references therein]{kilian2017structural}. These problems are exacerbated when the dimensionality of the system grows, making LPs our method of choice to obtain impulse responses in high dimensions.

We consider high-dimensional local projections (HDLPs) in a general time series framework where the number of regressors can grow faster than the sample size. Impulse response analysis quickly becomes high-dimensional in macroeconomic research. Even when considering impulse response analysis with few variables, the number of regressors in LPs or SVARs is often large due to the common practices of including many lags to control for autocorrelation (see e.g., \citealp{bernanke1998measuring}, \citealp{romer2004new} and \citealp{sims2006were}) or to robustify against (near) unit roots via lag augmentation as in \citet{olea2020local}. Similarly, additional regressors are often introduced to capture seasonal patterns in impulse responses (see e.g., quarter-dependent coefficients used in \citealp{blanchard2002empirical}), or to permit nonlinearities to produce state-dependent impulse responses (\citealp{koop1996impulse};  \citealp{ramey2018government}).

Additionally, one might be interested in impulse response analysis with many variables. Motivations of the latter include, amongst others, avoidance of contaminated measurements of policy innovations or informational deficiency with impulse response estimates that are distorted by omitted variable bias, the ability to use several observable measures of difficult-to-measure variables in theoretical models, or opportunities for researchers and policy makers to investigate the impact of shocks on a larger, possibly more disaggregated set of variables they care about (see e.g., \citealp{bernanke2005measuring}, \citealp{forni2014sufficient}, \citealp{stock2016DFM} and \citealp[Ch.~16]{kilian2017structural}).

Regular estimation approaches, however, become infeasible for high-dimensional settings. Traditional approaches to high-dimensionality include modelling commonalities between variables, such as through factor-augmented VARs (FAVAR) (\citealp{bernanke2005measuring}), dynamic factor models (DFM) (\citealp{forni2009opening}; \citealp{stock2016DFM}) or -- for panel structures -- global VARs (\citealp{ChudikPesaran16}); as well as Bayesian shrinkage methods (\citealp{banbura2010large}; \citealp{chan2020large}). Recently, sparse shrinkage methods such as the lasso have gained considerable popularity in econometrics for policy evaluation (\citealp{BCH14}) and (macroeconomic) time series analysis (\citealp{kock2020penalized}). Their use in impulse response analysis however has only scarcely been explored.

While several methods and theoretical results now exist for estimating sparse VAR models -- see e.g., \cite{BasuMichailidis15, KockCallot2015, masini2019regularized} and the references cited therein -- inference on impulse responses is complicated by two issues. First, sparse estimation techniques such as the lasso perform model selection, which induces issues with non-uniformity of limit results if this selection is ignored (\citealp{LeebPoetscher05}). Second, while several methods such as orthogonalization (\citealp{BCH14}) and debiasing, or desparsifying the lasso (\citealp{vandeGeer14}; \citealp{JavanmardMontanari14}) have been proposed to yield uniformly valid (or `honest') post-selection inference, the impulse response parameters are nonlinear functions of all estimated VAR parameters. This severely limits the applicability of existing post-selection inference methods which are typically designed for (relatively) low-dimensional parameters of interest that can be estimated directly. Indeed, to our knowledge, impulse response analysis in sparse HD-SVARs is only considered in \cite{krampe2022}, who construct a complex multi-step algorithm to overcome these complications. Instead, by casting the problem in the LP framework, we reduce the impulse response parameter(s) to a (directly estimable) low-dimensional object in the presence of high-dimensional nuisance parameters, which makes the standard post-selection tools available. 

We develop HDLP inference based on the desparsified lasso of \cite{vandeGeer14} and its time series extension in \cite{adamek2021lasso}. 
The latter has recently been empirically investigated in the context of LP estimation with instrumental variables (LP-IV) in contemporaneous work by \cite{Karapanagioti21}. Their approach not only differs in the focus on IV models, but they also apply the method of \cite{adamek2021lasso} `as is'. Instead, we tailor the approach of \cite{adamek2021lasso} specifically to high-dimensional LPs consisting of a small number of parameters of interest --  the dynamic response of a variable to a shock at a given horizon -- and many controls. In particular, we modify their approach by leaving the parameter of interest unpenalized during the estimation procedure to ensure it does not suffer from penalization bias. Such a setting is of more general relevance for treatment effect models consisting of a small number of variables whose effects are of interest combined with a large set of controls. We theoretically show that the combination of few unpenalized parameters with many penalized ones does not affect the asymptotic behaviour of the desparsified lasso.

Through simulation experiments, we show that our proposed estimator has considerably better coverage rates than the standard desparsified lasso in finite samples. Furthermore,  on an empirically calibrated DFM (\citealp{li2021local}) we find that our proposed HDLP method shows competitive performance in a sparse DFM compared to a low-dimensional LP and a factor-augmented LP.

We consider two canonical macroeconomic applications and demonstrate the performance of the proposed desparsified-lasso based estimator for HDLPs in recovering structural impulse responses. Specifically, we first extend the work of \citet{bernanke2005measuring} on macroeconomic responses to a shock in monetary policy to a HDLP setting.
Second, we consider the work by \citet{ramey2018government} on state-dependent impulse responses to a shock in government spending, as also studied by \cite{Karapanagioti21} in their LP-IV setting, but we extend the original specification used in \citet{ramey2018government} to a HDLP specification with more lags 
as well as a more complex state-dependent HDLP with interaction between different state variables. 
The methods needed for these applications are implemented in the R package \texttt{desla} (\citealp{desla}), which also offers the methods of \citet{adamek2021lasso}.

The paper is organized as follows. Section \ref{sec:HDLP} introduces the model and our inferential procedure,
as well as our main result on the asymptotic normality of the desparsified lasso.
Section \ref{sec:Simulations} contains simulation studies that examine the small sample performance of the proposed desparsified lasso. Section \ref{sec:empiricalapplications} presents the results of the two macroeconomic applications. Section \ref{sec:conclusion} concludes.

A word on notation. For any $N$ dimensional vector $\boldsymbol{x}$, $\left\Vert \boldsymbol{x}\right\Vert_r=\left(\sum_{i=1}^{N}\left\vert x_i\right\vert^r\right)^{1/r}$ 
denotes the $l_r$-norm, with the convention that $\norm{\boldsymbol{x}}_0 = \sum_{i} 1 (\abs{x_i}>0)$.  Depending on the context, $\sim$ denotes equivalence in order of magnitude of sequences, or equivalence in distribution. 

\section{High-dimensional Local Projections}\label{sec:HDLP}
Consider the local projection regression
\begin{equation}\label{eq:localprojection}
    y_{t+h}=\phi_{h}x_{t}+ \rho_{h}y_t  + \boldsymbol{\eta}_h^\prime\bw_{s,t}+\sum_{k=1}^K \boldsymbol{\delta}_{h,k}^\prime\bz_{t-k}+u_{h,t}, \quad h = 0, 1, \ldots, h_{\max},
\end{equation}
where $\phi_{h},  \rho_{h}, \boldsymbol{\eta}_h, \boldsymbol{\delta}_{h,k}$ are the projection parameters, $u_{h,t}$ is the projection error and the set of variables $\bz_t=\big(\bw_{s,t}^\prime,y_t,x_t, \bw_{f,t}^\prime\big)^\prime$ consist of the response $y_t$, the shock variable $x_t$,  and the vectors of control variables split into the ``slow" variables $\bw_{s,t}\in\mathds{R}^{n_s}$, and the ``fast" ones $\bw_{f,t}\in\mathds{R}^{n_f}$ for identification purposes, as discussed below.
\footnote{We omit the constant, since we generally demean the data when using the desparsified lasso.}
Our single parameter of interest is $\phi_{h}$, the dynamic response at horizon $h$ of $y_t$ after an impulse in $x_t$. The LP impulse response function of $y_t$ with respect to $x_t$ is given by $\{ \phi_{h}\}_{h\geq 0}$ and can be obtained by estimating  \cref{eq:localprojection} for 
$h=0,1,\dots,h_{\max}$. 

To permit causal interpretation of the impulse responses, structural assumptions about the data generating process are needed. Example \ref{ex:exampleSVMA} considers a Structural Vector Moving Average model (see Assumption 3 in \citealp{plagborg2021local}) as an example.

\begin{example}[SVMA]\label{ex:exampleSVMA}
\textnormal{Consider the Structural Vector Moving Average (SVMA)
\begin{equation}\label{eq:SVMA}
    {\bz_t}=\bmu+
    {\bA(L)}
    {\bepsilon_t},~~\bA(L)=\sum_{k=0}^\infty
    {\bA_{k}}L^k,~~\bepsilon_{t}\overset{i.i.d.}{\sim}N(\bzero,\bI),
\end{equation}
where $\bepsilon_t$ is a vector of structural shocks, $L$ is the lag operator, $\{\bA_{k}\}_k$ is  absolutely summable, and $\bA(x)$ has full row rank for all complex scalars $x$ on the unit circle. }
\end{example}

While an assumption such as \Cref{ex:exampleSVMA} is necessary to recover the structural impulse responses, it is not a sufficient condition: $\bA_0$ is not identified without further restrictions. 
We focus on impulse responses identified by a (partially) recursive structure. This can be implemented in LPs by partitioning the variables in $\bz_t$ into slow variables ($\bw_{s,t}$) and fast variables ($\bw_{f,t}$). The slow variables are predetermined with respect to $x_t$, while the fast variables can react contemporaneously to the shock variable $x_t$ and therefore only enter the equation with a lag. Equivalently, in the language of recursively identified SVARs, $\bw_{s,t}$ is ordered above $x_t$, while $\bw_{s,t}$ is ordered below. Note also that in \cref{eq:localprojection} we assume that $y_t$ is predetermined with respect to (ordered above) $x_t$; if the reverse were true one should remove $y_t$ from the right-hand side of \cref{eq:localprojection}.\footnote{In this situation, the model would be equivalent to equation (1) of \cite{plagborg2021local}.}

\Cref{eq:localprojection} can also be used outside of the recursive identification framework, for example when the structural shock is externally identified. Examples of the latter are the narrative monetary policy shock series of \cite{romer2004new}, or the military spending series of \cite{ramey2018government}. One can then directly take $x_t=\epsilon_{1,t}$ provided  the structural shock is observed without measurement error. Given the exogeneity (and thus predeterminedness) of such shocks, all other variables would then be considered to be ``fast'' and no variable would be included contemporaneously; see e.g. \Cref{ex:ramey}.

Section 3.2 of \cite{plagborg2021local} provides detailed discussions on how other identification schemes, such as sign restrictions, can be imposed while estimating \cref{eq:localprojection}; these arguments generally remain valid in our high-dimensional setup as well. However, identification through instrumental variables 
as in e.g.~\cite{stock2018identification}
becomes more complicated. Adapting IV estimation methods such as two-stage least-squares (\citealp[Section 3.3]{plagborg2021local}) to high dimensions requires a different setup for the desparsified lasso (see e.g.~\citealp{gold2020inference}), which has not yet been explored for time series models. This remains an interesting avenue for future research.

While the LP in \cref{eq:localprojection} has a single parameter of interest, it can easily be extended to allow for LPs with a small number of parameters of interest, see Example \ref{ex:ramey}.
\begin{example}[State-Dependent LP]\label{ex:ramey}
\textnormal{Consider the nonlinear state-dependent model of \cite{ramey2018government}:
\begin{equation}\label{eq:statedependent}
\small
    y_{t+h}= I_{t-1}\left[\alpha_h+\phi_{A,h}x_{t}+\sum_{k=1}^K\boldsymbol{\delta}_{A,h,k}^\prime\bz_{t-k} \right]
    +(1-I_{t-1})\left[\phi_{B,h}x_{t}+\sum_{k=1}^K\boldsymbol{\delta}_{B,h,k}^\prime\bz_{t-k} \right]+u_{h,t}.
\end{equation}
Here, $\phi_{A,h}$ and $\phi_{B,h}$ are the two responses of $y_t$ after an impulse in $x_t$, corresponding to two  different states distinguished by the dummy variable $I_t$.} 
\end{example}

To generally accommodate LP settings with a small number of parameters of interest and a large number of controls, we will use the more general notation
\begin{equation}\label{eq:DGP}
    {y_t}=\underset{1\times H}{\bx_{\cH,t}^\prime}\underset{H\times 1}{\bbeta_\cH}+\underset{1\times(N-H)}{\bx_{-\cH,t}^\prime}\underset{(N-H)\times 1}{\bbeta_{-\cH}}+u_t,~~t=1,\dots,T,
\end{equation}
and in matrix form
\begin{equation*}
\by=\bX_\cH\bbeta_\cH+\bX_{-\cH}\bbeta_{-\cH}+\bu.
\end{equation*}
We hereby separate the parameter vector $\bbeta=[\bbeta_\cH^{\prime},\bbeta_{-\cH}^{\prime}]^{\prime}$ into two groups.
The first group consists of the $H$-dimensional set of parameters of interest $\bbeta_\cH$, corresponding to the variables $\bx_{\cH,t}$, which will be unpenalized during the estimation procedure to ensure they are not affected by penalization bias. The second group consists of the large set of parameters $\bbeta_{-\cH}$, corresponding to control variables $\bx_{-\cH,t}$, which will be penalized during the estimation procedure. We hereby use $\cH$  to denote the set that indexes the $H$ variables of interest. Without loss of generality, we order these variables first in $\bX=[\bX_\cH,\bX_{-\cH}]$. Finally, since each horizon $h$ of the LPs is estimated separately, we suppress the dependence on $h$ in our notation. 

\subsection{Local Projection Estimation} \label{subsec:estimator}
We start by estimating \cref{eq:DGP} with the lasso to obtain an initial estimate of $\bbeta$, while leaving the parameters of interest $\bbeta_{\cH}$ unpenalized.
The first stage lasso  is defined as 
\begin{equation}\label{eq:problem}
     \hat\bbeta^{(L)}=[\hat{\bbeta}_\cH^{(L)\prime},\hat{\bbeta}_{-\cH}^{(L)\prime}]^\prime=\argmin_{\bbeta^*\in\mathds{R}^{N}}\norm{\by-\bX\bbeta^*}_2^2/T+2\lambda\norm{\Weight\bbeta^*}_1, 
\end{equation}
where $\Weight$ is an $N\times N$ diagonal matrix with $\Weight_{i,i}=0$ for $i\in \cH$ and 1 otherwise. 
The Frish-Waugh-Lovell-style result of \cite{yamada2017} ensures that the penalized and unpenalized estimates can be respectively obtained by
\begin{equation*}\begin{split}
    \hat{\bbeta}^{(L)}_{-\cH}&=\argmin_{\bbeta^*\in\mathds{R}^{(N-H)}}\norm{\bM\by-\bM\bX_{-\cH}\bbeta^*}_2^2/T+2\lambda\norm{\bbeta^*}_1,\\
    \hat\bbeta_\cH^{(L)}&=\hat\bSigma_\cH^{-1}\bX_\cH^\prime(\by-\bX_{-\cH}\hat\bbeta_{-\cH}^{(L)})/T,
\end{split}\end{equation*}
where the residual maker $\bM:=(I-\bX_\cH(\bX_\cH^\prime\bX_\cH)^{-1}\bX_\cH^\prime)$, and $\hat\bSigma_\cH:=\frac{\bX_\cH^\prime\bX_\cH}{T}$. 
We can therefore use regular lasso estimation on the transformed response $\bM\by$ and predictors $\bM\bX_{-\cH}$ to obtain $\hat{\bbeta}^{(L)}_{-\cH}$. 

We next ``desparsify'' this estimate to allow for uniformly valid inference. The desparsified lasso estimator is defined as 
\begin{equation}\label{eq:MDL}
    \hat\bbeta_\cH:=\hat\bbeta_\cH^{(L)}+\hat\bTheta\bX^\prime(\by-\bX\hat\bbeta^{(L)})/T,
\end{equation}
where $\hat\bTheta$ is an $H\times N$ submatrix of an approximate inverse of $\hat\bSigma:=\frac{\bX^\prime\bX}{T}$. We obtain $\hat\bTheta$ from `nodewise' regressions, which estimate the linear projections of $\bx_j$ onto $\bX_{-j}$, where $\bx_j$ is the $j$th column of $\bX$ and $\bX_{-j}$ is $\bX$ without the $j$th column. We let
\begin{equation*}
    \hat\bgamma_j=\argmin_{\bgamma_j^*\in\mathds{R}^{N-1}}\norm{\bx_j-\bX_{-j}\bgamma_j^*}_2^2/T+2\lambda_j\norm{\bgamma_j^*}_1, 
\end{equation*}
and construct $\hat\bTheta:=\hat\bUpsilon^{-2} \hat\bGamma$, where
$\hat\bUpsilon^{-2}:=\diag(1/\hat\tau^2_1,\dots 1/\hat\tau^2_H)$, 
with $\hat\tau_j^2=\norm{\hat\bv_j}_2^2/T+\lambda_j\norm{\hat\bgamma_j}_1, \hat\bv_j=\bx-\bX\hat\bgamma_j$
and $\hat\bGamma$ stacks the $\hat\bgamma_j$'s row-wise: 
\begin{equation*}
    \hat\bGamma:=\left[\begin{array}{ccccc} 
    1 &  \dots & -\hat\gamma_{1,H} & \dots & -\hat\gamma_{1,N}\\
    \vdots & \ddots & \vdots & & \vdots \\
    -\hat\gamma_{H,1} & \dots & 1 & \dots & -\hat\gamma_{H,N}
    \end{array}\right].
\end{equation*}

The specific structure of the LPs consisting of the same set of regressors at each horizon $h$ allows for
computational and efficiency gains in obtaining the desparsified lasso.
The initial lasso in  \cref{eq:problem} should be obtained for each horizon, but
the population linear projection coefficients $\bgamma_j$ in the nodewise regression 
do not change with the horizon. It is therefore sufficient to compute the nodewise lasso estimator once. This can be done best at horizon zero where most time points are available for estimating the nodewise regression.
We therefore partly avoid the loss of efficiency 
at further horizons where the most recent observation at each new horizon would be lost for estimation otherwise. As an example, 
for the LP in \cref{eq:localprojection}, we
estimate 
\begin{equation}\label{eq:nodewiseLP}
    x_t=\bpsi_0^\prime(y_t,{\bw}_{s,t}^\prime)^\prime+\sum_{k=1}^K\bpsi_k^\prime\bz_{t-k} +v_{1,t},
\end{equation}
then store $\hat\bTheta$ constructed from $\hat\bgamma_{1}=\left[\hat\bpsi_0^{\prime},\dots, \hat\bpsi_{K}^{\prime}\right]^\prime$, and the residual vector $\hat\bv_1$. The former
can then be re-used at each horizon in the 
desparsified lasso estimator in \cref{eq:MDL}, while
the latter can be re-used in the estimation of the
long-run covariance matrix which enters the asymptotic distribution of the desparsified lasso, see Section \ref{subsec:inference}.

Finally, regarding the choice of tuning parameter $\lambda$ in the initial regression and $\lambda_j$ in the nodewise regressions, 
we follow the approach of \citet{adamek2021lasso} who propose an iterative plug-in procedure that simulates the quantiles of an appropriate ``empirical process'' which should be dominated by $\lambda$; see their Section 5.1 for details. 

\subsection{Local Projection Inference} \label{subsec:inference}
We next discuss the asymptotic properties of 
$\hat\bbeta_{\cH}$. The assumptions needed for our analysis are listed in detail in Appendix A, and are discussed briefly below. Detailed discussions can be found in \cite{adamek2021lasso}, on whose approach our theory is built.

We need assumptions on the DGP (\cref{ass:DGP}), the sparsity of the parameter (\ref{ass:Sparsity}), regularity conditions for high-dimensional settings (\ref{ass:Eigenvalue}) and assumptions on the set of parameters on which inference is conducted (\ref{ass:HFinite}).
\cref{ass:DGP}\ref{ass:DGPStationary} requires that 
the regressors are contemporaneously uncorrelated with the error term, and that the error terms 
and the variables in $\bz_{t}$ have finite unconditional moments up to a certain order. Crucially, \cref{ass:DGP}\ref{ass:DGPStationary} allows for serial correlation in the error terms, which is a typical feature of LP regressions.
\cref{ass:DGP}\ref{ass:DGPNED} assumes $\bz_{t}$ to be near-epoch-dependent (NED). An NED process can be interpreted as a process that is well-approximated by a mixing process, but does not have to be mixing itself. Unlike mixing conditions, which may exclude even very simple autoregressive processes, NED conditions permit very general forms of dependence often encountered in macroeconomic research, see 
\cite{adamek2021lasso} for a more detailed discussion.

Comparing our DGP assumptions 
to the SVMA assumptions of Example \ref{ex:exampleSVMA}, 
one can verify that the moment condition is satisfied trivially under Gaussianity, and the NED assumption can be shown to hold when the VMA coefficients $\bA_k$ converge to 0 sufficiently quickly; see Example 17.3 of \cite{Davidson02}. 
If one would relax the Gaussianity assumption, a trade-off would occur between the number of moments and the allowed dependence -- the more moments, the more persistent the process may be.

Second, we impose a sparsity assumption on the parameter $\bbeta$ (\cref{ass:Sparsity}), as typically required for lasso estimators.
We derive our results under weak sparsity, thereby recognizing that the true parameters are likely not exactly zero. In particular, we assume
that $\norm{\bbeta}_{r}^r$ is sufficiently small for some $0\leq r<1$. 
While for $r=0$ our results boil down to 
requiring exact sparsity 
for larger values of $r$, we allow for many non-zero entries in $\bbeta$ provided they are relatively small. The smaller $r$, the more restrictive this assumption therefore is. We also require weak sparsity in the nodewise regressions, which
amounts to assuming that the inverse population covariance matrix $\bSigma^{-1}:=\left(\E\bX^\prime\bX/T\right)^{-1}$ is sparse in (induced) $r$-norm. Since \cref{eq:localprojection} has a similar regressor structure as a VAR, this can be shown to hold under conditions similar to the sparse VAR in Example 6 of \cite{adamek2021lasso}.

Next, we require  that the minimum eigenvalue of $\bSigma$ is bounded away from 0 (\cref{ass:Eigenvalue}).
This assumption ensures that the population covariance matrix of the regressors remains well-behaved with increasing $N$. Again, thanks to the similarity of  \cref{eq:localprojection} to a VAR, lower bounds on the minimum eigenvalue can be derived using the results on page 6 of \citet{masini2019regularized}. Finally, we also require that the number of unpenalized parameters of interest to be bounded (\cref{ass:HFinite}). In LPs, this is trivially satisfied, for instance $H=1$ in \cref{eq:localprojection}, and $H=3$ in \cref{eq:statedependent}.\footnote{While the impulse responses are the two parameters of interest, we may also want to leave the dummy parameter unpenalized.} 

Under these assumptions, \cref{thm:Inference} establishes the asymptotic normality of our desparsified lasso estimator, which allows for valid asymptotic inference. 

\begin{theorem}\label{thm:Inference}
Under \cref{ass:DGP,ass:Sparsity,ass:Eigenvalue,ass:HFinite,ass:asymptoticrates}, for the model in \cref{eq:DGP}, we have
\begin{equation*}
\sup_{\underset{z\in\mathds{R}} {\boldsymbol{\beta} \in \boldsymbol{B}_N(r,s_r)}} \left\vert\P\left(\sqrt{T} \frac{\br' (\hat\bbeta_\cH-\bbeta_\cH)}{\sqrt{\br' \hat\bUpsilon^{-2} \hat{\bOmega} \hat{\bUpsilon}^{-2} \br}} \leq z\right)-\Phi(z)\right\vert=o_p(1),
\end{equation*}
where $\Phi(\cdot)$ is the CDF of $N(0,1)$, $\br\in\mathds{R}^\cH$ is chosen to test hypotheses of the form $\br^\prime \bbeta_H=0$, and $\hat{\bOmega}$ is a consistent long-run covariance estimator.
\end{theorem}

Note that in LPs, the errors $u_{h,t}$ from \cref{eq:localprojection} are generally autocorrelated due to the leads of $y_t$ on the left-hand side. As in \cite{adamek2021lasso}, we therefore use an autocorrelation robust Newey-West long-run covariance estimator
$\hat\bOmega:=\sum_{l=1-Q_T}^{Q_T-1}\left(1-l/Q_T\right)\hat\bXi(l)$,
with bandwidth $Q_T<T$, where 
$\hat\bXi(l):=\frac{1}{T-l}\left.\sum_{t=l+1}^{T}\hat\bq_t\hat\bq_{t-l}^{\prime}\right.$, $\hat\bq_t:=\left(\hat v_{1,t}, \dots, \hat v_{H,t}\right)^\prime\hat u_t$. 
To obtain our asymptotic results, we generally require that the bandwidth $Q_T$ grows as $T\to\infty$ at a sufficiently slow rate, see \cref{ass:asymptoticrates} in Appendix A for details. Regarding the choice of $Q_T$ in finite samples, we use the bandwidth estimator of  \cite{andrews1991heteroskedasticity}. 

For the LP in \cref{eq:localprojection}, Theorem \ref{thm:Inference}
permits asymptotic confidence intervals 
$CI_h(\alpha):=\left[\hat\phi_h\pm z_{\alpha/2}\sqrt{T^{-1}\hat{\omega}_{ h}\hat{\tau}_{h}^{-4}}\right]$,
where $\hat\omega_{h}$ and $\hat\tau_{h}^{-2}$ are the scalar versions of $\hat\bOmega_h$ and $\hat\bUpsilon^{-2}_h$ respectively, and $z_{\alpha/2}:=\Phi^{-1}(1-\alpha/2)$. Note the dependence on $h$ in this result, since the residuals $\hat \bu=\hat \bu_{h}$ are different at each horizon, and $\hat\bOmega=\hat\bOmega_h$ and $\hat\bUpsilon^{-2}=\hat\bUpsilon_h^{-2}$ as well.

\section{Simulations}\label{sec:Simulations}
We perform a simulation study to analyse the finite sample performance of the desparsified lasso estimator.
In \cref{sec:Partial_penalization} we compare our proposed method with unpenalized parameter of interest to the standard desparsified lasso in a sparse structural VAR. In \cref{sec:DFM}, we study our proposed method in an empirically calibrated DFM.

\subsection{Sparse Structural VAR Model}\label{sec:Partial_penalization}
Consider the structural VAR with structural shocks $\bepsilon_t$
\begin{equation}\label{eq:sim}
    \left(\begin{array}{c}
        y_t\\
        \bw_{s,t}  
    \end{array}\right)=\underset{ P\times 1}{\bz_t}=\sum_{k=1}^4\bA_{k}\bz_{t-k}+\bepsilon_t,\quad \bepsilon_t\overset{iid}{\sim}N(\bzero,\bI),
\end{equation}
where the autoregressive parameter matrices $\bA_k$ are tapered Toeplitz matrices.
We analyse the response of $y_t$ to the first shock $\epsilon_{1,t}$ and investigate the coverage and width of 95\% point-wise confidence intervals corresponding to the impulse response parameters at horizons 1 to 10.
We consider different values for the number of variables $P=\{20, 40, 100\}$ and time series length $T =\{100, 200, 500\}$ and report results for 1000 replications. Details on the DGP and implementation of the HDLP are included in Appendix \ref{app:sparse_var}. %Appendix C.1.1.

\begin{figure}[t]
\centering
\includegraphics[width=\textwidth]{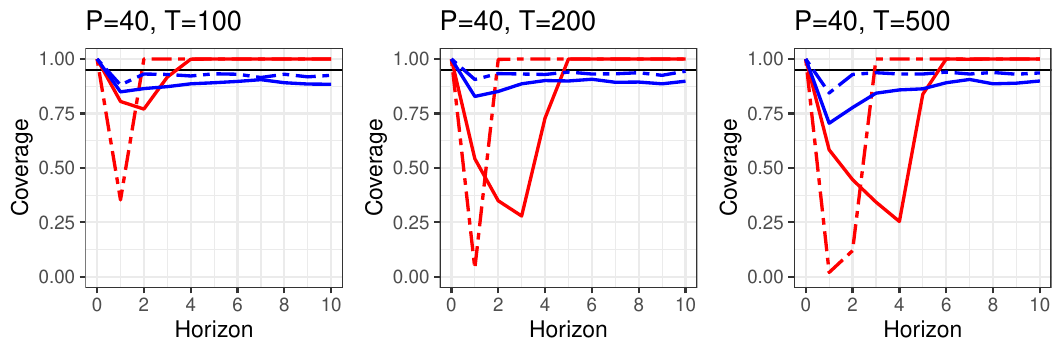}
\caption{Coverage rates of the standard desparsified lasso (red) and the proposed desparsified lasso with $\phi_{h}$ unpenalized (blue). Dashed lines indicate results for the second 
DGP. }\label{fig:simresults}
\end{figure}

\Cref{fig:simresults} presents the coverage rates for the proposed desparsified lasso (blue), which leaves the impulse response parameters of interest unpenalized, in comparison to the standard desparsified lasso (red). We only display  results for $P=40$, but the results for $P=20
$ and $P=100$ are very similar and included in \Cref{fig:S1} %Figure C.1 
 in Appendix \ref{app:extra_figures}, 
 alongside the interval widths in \Cref{fig:S2}. %Figure C.2.

Our proposed estimator attains good coverage rates close to the nominal level of 95\% for all combinations of $P$ and $T$. 
For both DGPs, coverage rates are usually the lowest around 70-90\% for horizon 1, but revert to the nominal level for larger horizons. Good coverage at further horizons is to be expected, since the local projection coefficients will tend to become more sparse in stationary models.
Importantly, considerable gains are obtained compared to the standard desparsified lasso. 
Especially at shorter horizons -- which are usually of most interest in practice -- coverage rates of the latter drop to 25\% in the first DGP and even more  in the second. We also see in \Cref{fig:S2} %Figure C.2 
(Appendix \ref{app:extra_figures}) 
that the interval widths are broadly similar or even smaller than the ones for the standard desparsified lasso, meaning we obtain better coverage without losing power. 

The poor coverage rates of the standard desparsified lasso occur since the parameters of interest are not selected in the initial lasso regression.
A simple alteration of the desparsified lasso that leaves this parameter unpenalized thus brings the coverage rates much closer to the nominal level. Note that the standard desparsified lasso has coverage exceeding our proposed estimator at further horizons; this is because the true impulse response becomes close to zero, where the lasso's bias towards zero is beneficial. This does not run counter to our conclusion, as we believe more uniform coverage over different parameter values is desirable.

\subsection{Empirically Calibrated Dynamic Factor Model}\label{sec:DFM}
We now examine the performance of our proposed estimator on an empirically calibrated dynamic factor model, inspired by \cite{lazarus2018har} and \cite{li2021local}. 
Unlike the former works which use the quarterly dataset of \cite{stock2016DFM} to fit their DFM, we use
the monthly FRED-MD database (\citealp{mccracken2016fred})  which we analyse 
in \Cref{sec:empiricalapplications}.
Implementation details on the simulation set-up are in Appendix \ref{sec:DFM_details}, %Supplement S1.2,
an overview on the variables in the FRED-MD database  in Appendix \ref{sec:data}. %Supplement S2.

We consider both a ``dense'' DFM where the factor loadings are obtained using principal components and a factor VAR is estimated by OLS, and a ``sparse'' DFM where the loadings are obtained using the WF-SOFAR estimator of \cite{uematsu2022estimation} and the VAR is estimated by the lasso.
The dense DFM is inherently difficult for our proposed estimator since the covariance structure of the variables is dense, whereas we assume it to be sparse in \Cref{ass:Sparsity}, and the DGP is likely to violate the sparsity of the nodewise regression coefficients.
However, bounds on the sparsity of the nodewise regressions can be obtained for sparse factor models and VARs, see Example 5 and 6 of \cite{adamek2021lasso}. 
We therefore expect
better coverage properties of our proposed estimator is this setting. 
The
WF-SOFAR estimator seems particularly appropriate to use since \cite{uematsu2022inference} find evidence of a sparse factor structure in the FRED-MD data.\footnote{The difference in sparsity between the dense and sparse DFMs is relatively limited. For details, see \Cref{fig:S3} in Appendix \ref{sec:DFM_details}.}

\begin{figure}[t]
\centering
\includegraphics[width=\textwidth]{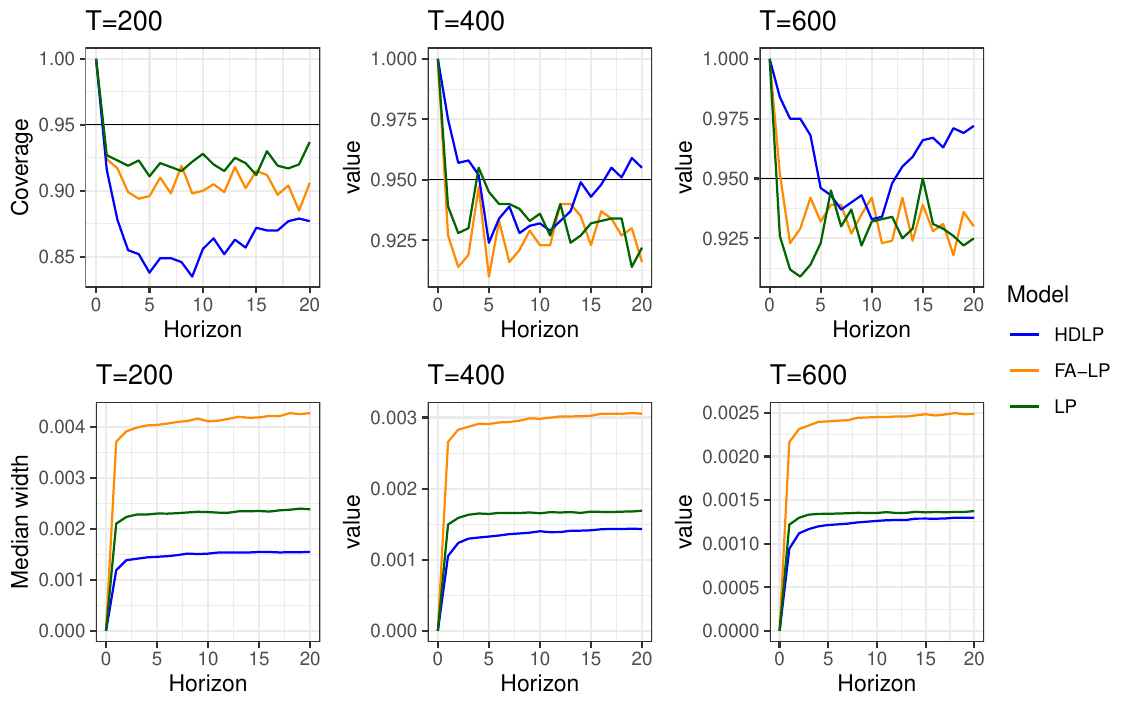}
\caption{Coverage rates and median interval widths
in the sparse DFM.}\label{fig:sparse_models}
\end{figure}

We study the impulse response of Industrial Production (IP) to the Federal Funds Rate (FFR). 
Our HDLP method estimates \cref{eq:localprojection} with $y_t=\text{IP}_t$, $x_t=\text{FFR}_t$, the remaining 120 variables in $\bw_{s,t}$,  $\bw_{f,t}$ empty, and 3 lags included.\footnote{Each local projection therefore has 488 regressors.}
We compare the performance of our proposal to two benchmarks. 
The first benchmark is a factor-augmented local projection, labelled ``FA-LP''. 
We favour FA-LP by including the true number of six factors, estimated by principal components from the 120 variables, and estimate the 
FA-LP by OLS. 
As this method matches the true DGP closely, we expect this benchmark to be a highly competitive standard. 
Our second benchmark 
is a 3-variable LP (estimated by OLS) with consumer price index as third variable as in \cite{bernanke2005measuring}. This benchmark, labelled ``LP", allows us to assess the gain in using high-dimensional data. 
All models are estimated using the R package \texttt{desla} (\citealp{desla}), taking the plug-in constant equal to 0.4 for the HDLP.

In the dense DFM, see \Cref{fig:S4} %Figure S.4
in Appendix C, our proposed HDLP method attains  
coverage around 70\% for the first 5 periods, then steadily improves at further horizons where it reaches nominal coverage around horizon 20 for large sample sizes. 
The benchmarks perform, however, considerably better as expected.
They perform similarly at $T=200$ 
with 90\% coverage, but diverge at larger sample sizes. The FA-LP model has improved coverage at larger samples, reaching nominal at $T=600$, whereas the coverage of the LP deteriorates at larger samples.  

In the sparse DFM (\Cref{fig:sparse_models}), the performance of the HDLP considerably improves. Its coverage (top panels) is still below the benchmarks for $T=200$, stays around 85\% for all horizons, but it matches or exceeds the benchmarks at larger sample sizes. The performance of the FA-LP is similar to that of the dense DFM, and the LP benchmark maintains similar coverage to the FA-LP even at larger sample sizes; the 3-variable LP is likely a better approximation of the sparse DFM than the dense DFM.
Importantly though, we see across all sample sizes that the widths (bottom panels) of the FA-LP benchmark are 2-4 times larger than those of the HDLP, while the LP widths are also wider everywhere though to a lesser extent.

One should keep in mind that we set up the simulations advantageously to the FA-LP by using the true number of factors. The superior performance of the FA-LP, particularly for the dense model where our HDLP is expected to suffer, is thus hardly surprising. 
Yet for the sparse DFM, our HDLP method outperforms both the idealized FA-LP and the 3-variable LP 
with comparable or superior coverage 
yet considerably tighter intervals.

To conclude our simulation experiments, 
we investigate the sensitivity of our method to the choice of long-run covariance estimator. We compare the Newey-West (NW) estimator to 
the fixed-$b$ Newey-West 
and  Equal-Weighted Cosine (EWC) estimator 
suggested by \cite{lazarus2018har}.
In Appendix \ref{sec:other_LRVs} %Supplement S1.3 
we find that the first generally performs slightly worse than our default choice of Newey-West (likely due to our adaptive choice of bandwidth), whereas the second performs very well in terms of coverage but has confidence intervals around 4-6 times as wide as the ones from our NW estimator. In settings where our method has good coverage, such as the sparse DFM, EWC  thus provides little benefit.
We therefore proceed with our default choice of NW estimator in the remainder of the paper, but these alternatives are available in the \texttt{desla} package.

\section{Structural Impulse Responses Estimated by HDLPs}\label{sec:empiricalapplications}
We apply the desparsified lasso for HDLPs to two canonical macroeconomic applications. 
In \Cref{subsec:bernanke}, we build on the work by \cite{bernanke2005measuring} on monetary policy analysis,
in 
\Cref{subsec:ramey} we build on the work by \cite{ramey2018government} on government spending.
All analyses are performed in \texttt{R} 
using the package
\href{https://CRAN.R-project.org/package=desla}{\texttt{desla}}.

\subsection{Impulse Responses to a Shock in Monetary Policy} \label{subsec:bernanke}
We consider the work by \citet{bernanke2005measuring}, wherein they estimate impulse responses to a monetary policy shock identified by the federal funds rate in a factor-augmented SVAR. 
We use HDLPs to perform the impulse response analysis and expand the dataset of the original paper
to the more recent and larger FRED-MD database \citep{McCrackenNg16}, thereby considering monthly data from January 1960 to 
October 2008
for 122 macroeconomic variables. For a full description of the data, see Appendix \ref{sec:data}.

We estimate the HDLP in \cref{eq:localprojection}, where as response variables $y_t$ we take the federal funds rate (FFR), industrial production (IP) and the consumer price index (CPI). As shock variable of interest $x_t$ we consider the federal funds rate (FFR). Furthermore, we use the identification strategy of \citet{bernanke2005measuring}, which consists of classifying variables as ``slow" in $\bw_{s,t}$ or ``fast" in $\bw_{f,t}$, according to whether these variables can respond within the same month to an unexpected change in the FFR.\footnote{As discussed before, this strategy is equivalent to recursive (partial) identification in an SVAR, as shown in Example 1 of \citet{plagborg2021local}. In fact, this application can also be seen as a high-dimensional extension of \citet{christiano2005nominal}.} 
For variables in FRED-MD that can be directly matched to their counterparts in the DRI/McGraw Hill Basic Economics Database, we use the same slow/fast classification as \citet{bernanke2005measuring}.
For new variables in FRED-MD, we apply the same general classification rule; i.e., variables in the categories Prices, Output \& Income, Labour Market, and Consumption are classified as slow, whereas those in the categories Interest \& Exchange Rates, Money \& Credit, Stock Market, and Housing as fast.\footnote{One exception in the Category Prices is the series ``OILPRICEx", which we classify as fast.} We apply the transformation codes of \citet{bernanke2005measuring} and \citet{McCrackenNg16} (for new variables) to remove stochastic trends. Using $K=13$ lags, the resulting HDLP consists of $N=1654$ regressors, while the time series length is $T=572$.\footnote{The exact number of variables in the model depends on which variable is taken as the response, and the effective number of observations depends on the horizon. These numbers are for the model where FFR is the response, at horizon 1.} 

\begin{figure}[t]
\centering
\includegraphics[width=\textwidth]{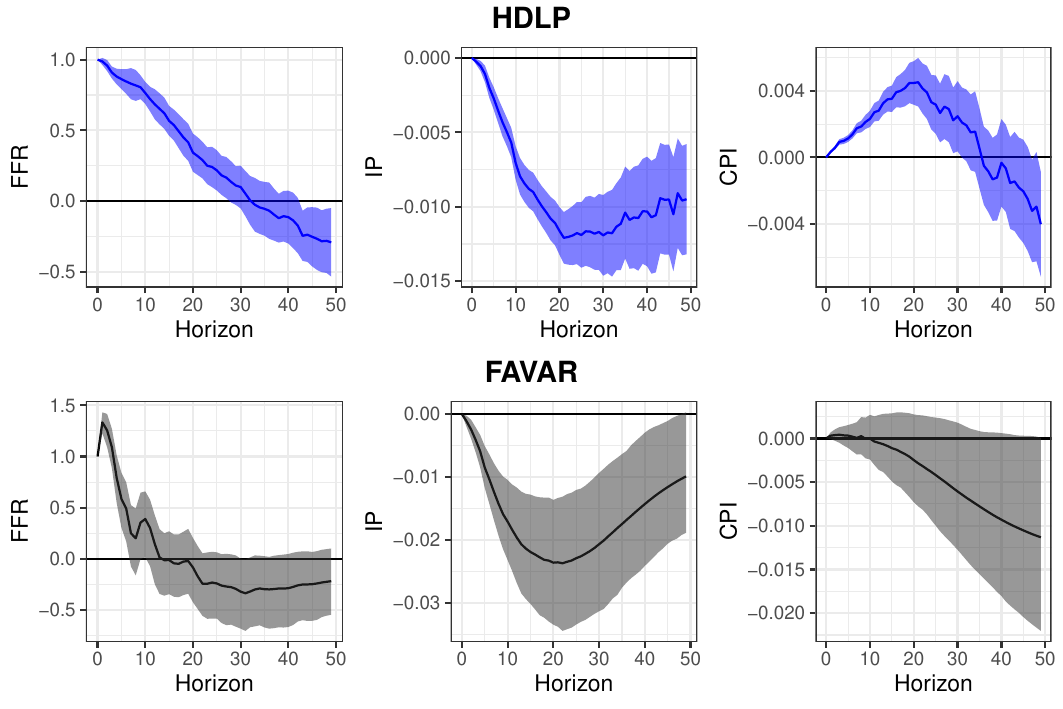}
\caption{Estimated impulse responses with 95\% confidence intervals of FFR, IP, and CPI to a monetary policy shock identified by the FFR for the HDLP and FAVAR. 
}\label{fig:FAVAR v HDLP}
\end{figure}

We compare the HDLP impulse responses obtained from the desparsified lasso to the ones obtained from a 3-factor FAVAR as used in \citet{bernanke2005measuring}.  Details about the FAVAR estimation are provided in Appendix \ref{app:FAVAR}. %Supplment S3. 
\Cref{fig:FAVAR v HDLP} shows the impulse responses of FFR, IP, and CPI to a shock in the FFR of a size such that the FFR has unit response at horizon 0.
Since we use the first difference of IP and CPI in the model, we cumulate the impulse response which then corresponds to the response in the original (level) variables.\footnote{Cumulative impulse response functions for the HDLP are obtained by cumulating the dependent variable, i.e. taking $\sum_{\ell=0}^{h}y_{t+\ell}$ on the LHS of \cref{eq:localprojection}. 
For the FAVAR, we take the cumulative sum of the regular impulse responses from previous horizons. To ensure that these impulse responses are comparable, we scale the FAVAR impulse responses by the response of the FFR at horizon 0. 
} The comparable figure in \citet{bernanke2005measuring} is Figure 1. 

The impulse responses from the FAVAR (bottom panel  \Cref{fig:FAVAR v HDLP}) and 
HDLP (top panel) are generally similar.
The impulse response of the FFR peaks at horizon 1 and then steadily declines to zero, which is in line with Figure 1 in \citet{bernanke2005measuring}. 
For the HDLP, the response stays at a higher level for a longer period than for the FAVAR. The response of IP is also in line with \citet{bernanke2005measuring}, with the largest drop around horizon 20, before eventually returning to zero. Notably, the response obtained with the HDLP is considerably smaller than the one for the FAVAR. Finally for the response of CPI, 
we see a more pronounced price puzzle with the HDLP. The response is positive and significant for 30 months, peaking at horizon 20. The FAVAR impulse response is largely in line with \citet{bernanke2005measuring}, with a small positive effect at early horizons, followed by a negative, though mainly insignificant, response after horizon 10. 

\subsection{Impulse Responses to a Shock in  Government Spending
} \label{subsec:ramey}
\cite{ramey2018government} estimate impulse responses to a shock in US government spending identified based on military spending news. In this paper, we first augment the authors' main LP specification with more lags as a robustness check, before considering an extended state-dependent HDLP specification with interacting states related to unemployment and recession conditions. We use the quarterly data 
provided by the authors at \href{https://econweb.ucsd.edu/~vramey/research.html#govt}{\texttt{econweb.ucsd.edu/$\sim$vramey}} covering the period 1889Q1 to 2015Q4. 

We estimate the state-dependent HDLP in \cref{ex:ramey} 
for 
real per capita GDP and government spending as $y_t$, while the shock variable $x_t$ is the military spending news shock.
We include taxes as controls in $\bw_{f,t}$, while no variables are included in $\bw_{s,t}$.\footnote{While \cite{ramey2018government} do not include taxes in their main analysis, they mention in footnote 11 that their results were robust to their inclusion. We prefer to include them as additional controls, following the VAR of \citet{blanchard2002empirical}.}
The state dummy variable $I_t$ distinguishes 
between low and high unemployment states, with $I_t=1$ when unemployment in period $t$ is larger than 6.5\%. 
For full details on the construction of the shock variable and the treatment of the other variables, see section II.B of \cite{ramey2018government}. We use $K=40$ lags, such that the resulting HDLP consists of $N=464$ regressors, while the time series length is $T=161$.\footnote{As in \Cref{subsec:bernanke}, these correspond to the model estimated at horizon 1.} 

\begin{figure}[t]
\centering
\includegraphics[width=\textwidth]{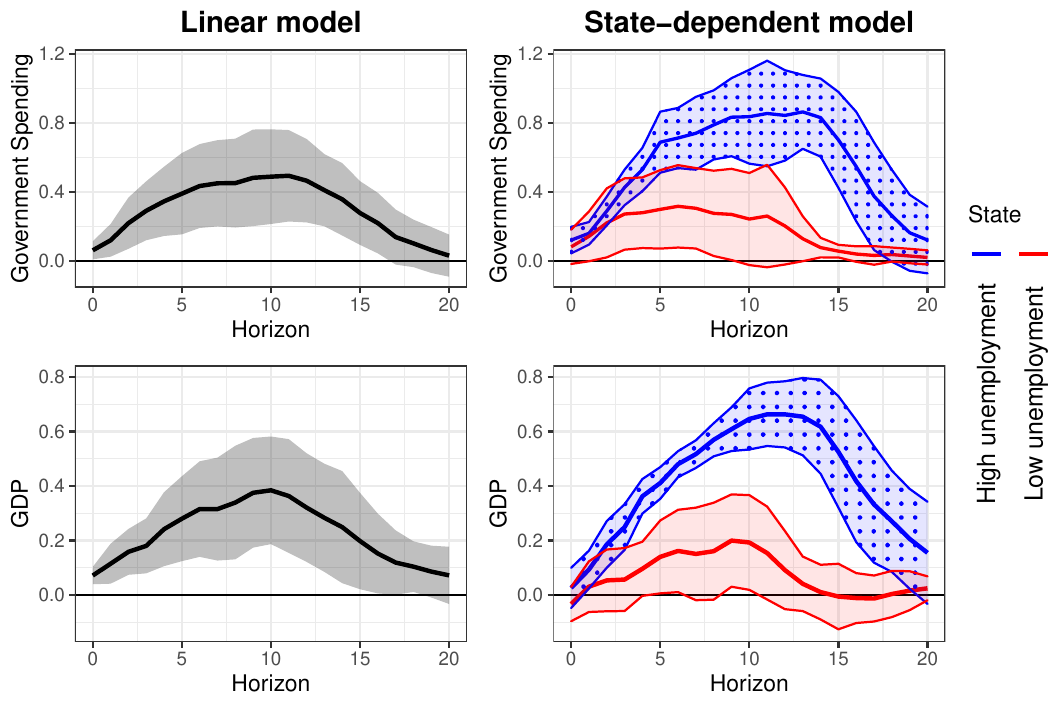}
\caption{Estimated impulse responses with 95\% confidence intervals of government spending and GDP to a government spending shock, in a model with 40 lags.}\label{fig:overall_RZ_40lags} 
\end{figure}

Figure \ref{fig:overall_RZ_40lags} shows the impulse responses to a shock in government spending. Note that the military spending news variable is scaled by GDP such that the impulse responses can be seen as a reaction to a shock of size 1\% of GDP. The comparable figure in \cite{ramey2018government} is Figure 5. The general shape and magnitude of these impulse responses are similar to the LP model with only 4 lags used in \cite{ramey2018government}. In our analysis, the peak of the responses in the high unemployment state occurs 2-3 quarters earlier, and experiences a sharper drop after horizon 15. We also find that the impulse response is not significant at horizon 0 in the linear and low-unemployment state.

\begin{figure}[t]
\centering
\includegraphics[width=\textwidth]{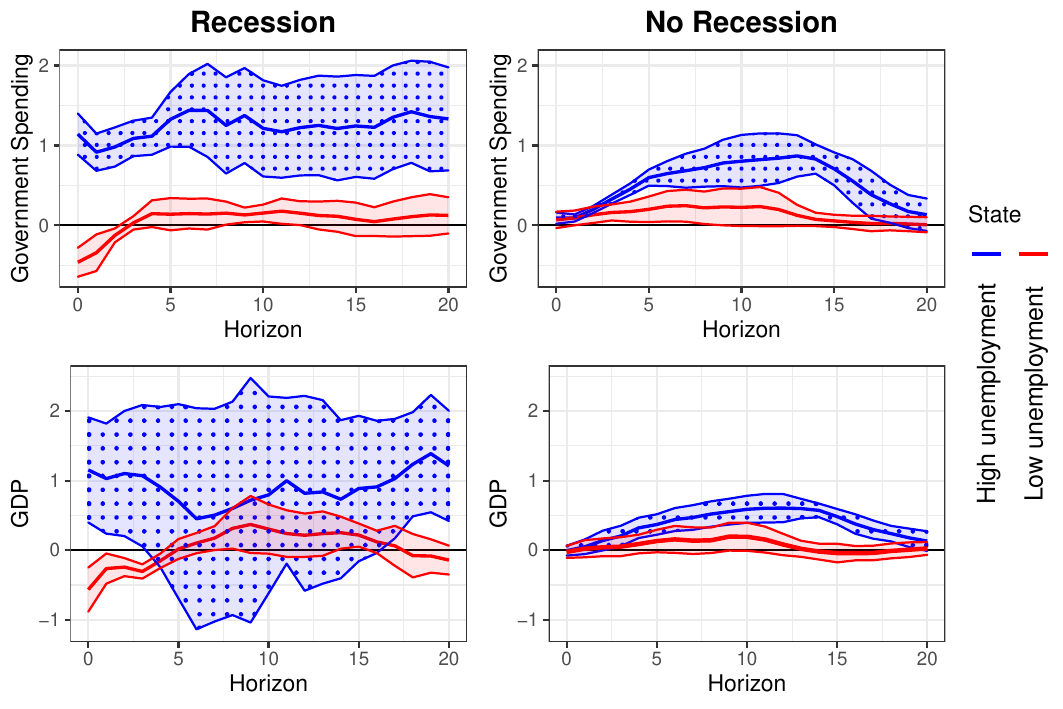}
\caption{Estimated Impulse responses with 95\% confidence intervals of government spending and GDP to a government spending shock.}\label{fig:2_states}
\end{figure}

Next to allowing the inclusion of more lags, the HDLP framework can  easily be extended to allow for $S$ states. 
The LP is then given by
\begin{equation*}
y_{t+h}=\sum_{i=1}^S \alpha_i I_{i,t-1} + \sum_{i=1}^S I_{i,t-1} \left[\phi_{i,h}x_{t}+\sum_{k=1}^K\boldsymbol{\delta}_{i,h,k}^\prime \bz_{t-k}\right]+u_{h,t}.
\end{equation*}
In addition to the unemployment state dummy $I_t^{(U)}$, we consider recession dummy $I_t^{(R)}$.\footnote{As in \cite{ramey2018government}, we follow the NBER classification of recession periods.}
Letting them interact
results in $S=4$ distinct state dummies.

\Cref{fig:2_states} displays the impulse response functions with four states. The impulse responses out of recessions resemble the state-dependent model of \Cref{fig:overall_RZ_40lags}, both in shape and magnitude, but we see a different pattern in the recession state. During recessions with low unemployment, there is a negative impact response for both government spending and GDP, which becomes insignificant after a few periods. In the high unemployment state, in contrast, we see a large, positive, and persistent effect on both variables.

\section{Conclusion} \label{sec:conclusion}
In this paper, we propose a modified version of the desparsified lasso to estimate local projections in high dimensions. 
The modification is simple, namely leaving the (small number of) dynamic impulse response parameter(s) of interest unpenalized in the initial lasso regression, yet provides considerable improvements in finite sample performance in terms of better coverage. The modified desparsified lasso estimator still comes with desirable asymptotic normality and uniformly valid asymptotic inference. Finally, we show how this method performs in a simulation using an empirically calibrated dynamic factor model and in canonical macroeconomic applications where we use the high-dimensional local projections framework to estimate structural impulse responses.

\section*{Acknowledgements}
We thank the 
co-editor and two referees for their thorough review 
which substantially improved the quality of the manuscript.
The first and second author were financially supported by the Dutch Research Council (NWO) under grant number 452-17-010.
Previous versions of this paper were presented at seminars at Aarhus University and ESSEC Business School, the NESG 2022 conference and the 2022 Maastricht Workshop on Dimensionality Reduction and Inference in High-Dimensional Time Series. We gratefully acknowledge comments by participants at these seminars and conferences. We also thank Otilia Boldea and Lenard Lieb for helpful discussions. Remaining errors are our own.

%%% REFERENCES
\bibliographystyle{chicago}
\bibliography{bibliography}

\begin{thebibliography}{}

\bibitem[\protect\citeauthoryear{Adamek, Smeekes, and Wilms}{Adamek et~al.}{2022a}]{desla}
Adamek, R., S.~Smeekes, and I.~Wilms (2022a).
\newblock {\em {desla: Desparsified Lasso Inference for Time Series}}.
\newblock \texttt{R} package version 0.2.0.

\bibitem[\protect\citeauthoryear{Adamek, Smeekes, and Wilms}{Adamek et~al.}{2022b}]{adamek2021lasso}
Adamek, R., S.~Smeekes, and I.~Wilms (2022b).
\newblock Lasso inference for high-dimensional time series.
\newblock {\em Journal of Econometrics\/}, Forthcoming.

\bibitem[\protect\citeauthoryear{Andrews}{Andrews}{1991}]{andrews1991heteroskedasticity}
Andrews, D.~W. (1991).
\newblock {Heteroskedasticity and autocorrelation consistent covariance matrix estimation}.
\newblock {\em Econometrica\/}~{\em 59}, 817--858.

\bibitem[\protect\citeauthoryear{Angrist, Òscar Jordà, and Kuersteiner}{Angrist et~al.}{2018}]{angrist2018semiparametric}
Angrist, J.~D., Òscar Jordà, and G.~M. Kuersteiner (2018).
\newblock {Semiparametric estimates of monetary policy effects: string theory revisited}.
\newblock {\em Journal of Business \& Economic Statistics\/}~{\em 36}, 371--387.

\bibitem[\protect\citeauthoryear{Ba{\'n}bura, Giannone, and Reichlin}{Ba{\'n}bura et~al.}{2010}]{banbura2010large}
Ba{\'n}bura, M., D.~Giannone, and L.~Reichlin (2010).
\newblock Large {B}ayesian vector auto regressions.
\newblock {\em Journal of Applied Econometrics\/}~{\em 25}, 71--92.

\bibitem[\protect\citeauthoryear{Basu and Michailidis}{Basu and Michailidis}{2015}]{BasuMichailidis15}
Basu, S. and G.~Michailidis (2015).
\newblock Regularized estimation in sparse high-dimensional time series models.
\newblock {\em Annals of Statistics\/}~{\em 43\/}(4), 1535--1567.

\bibitem[\protect\citeauthoryear{Belloni, Chernozhukov, and Hansen}{Belloni et~al.}{2014}]{BCH14}
Belloni, A., V.~Chernozhukov, and C.~Hansen (2014).
\newblock Inference on treatment effects after selection among high-dimensional controls.
\newblock {\em Review of Economic Studies\/}~{\em 81}, 608--650.

\bibitem[\protect\citeauthoryear{Bernanke, Boivin, and Eliasz}{Bernanke et~al.}{2005}]{bernanke2005measuring}
Bernanke, B.~S., J.~Boivin, and P.~Eliasz (2005).
\newblock Measuring the effects of monetary policy: a factor-augmented vector autoregressive ({FAVAR}) approach.
\newblock {\em Quarterly Journal of Economics\/}~{\em 120}, 387--422.

\bibitem[\protect\citeauthoryear{Bernanke and Mihov}{Bernanke and Mihov}{1998}]{bernanke1998measuring}
Bernanke, B.~S. and I.~Mihov (1998).
\newblock {Measuring monetary policy}.
\newblock {\em Quarterly Journal of Economics\/}~{\em 113}, 869--902.

\bibitem[\protect\citeauthoryear{Blanchard and Perotti}{Blanchard and Perotti}{2002}]{blanchard2002empirical}
Blanchard, O. and R.~Perotti (2002).
\newblock {An empirical characterization of the dynamic effects of changes in government spending and taxes on output}.
\newblock {\em Quarterly Journal of Economics\/}~{\em 117}, 1329--1368.

\bibitem[\protect\citeauthoryear{B{\"u}hlmann and van De~Geer}{B{\"u}hlmann and van De~Geer}{2011}]{SHD11}
B{\"u}hlmann, P. and S.~van De~Geer (2011).
\newblock {\em Statistics for High-Dimensional Data: Methods, Theory and Applications}.
\newblock Springer.

\bibitem[\protect\citeauthoryear{Chan}{Chan}{2020}]{chan2020large}
Chan, J.~C. (2020).
\newblock Large {B}ayesian vector autoregressions.
\newblock In P.~Fuleky (Ed.), {\em Macroeconomic Forecasting in the Era of Big Data}, Volume~52 of {\em Advanced Studies in Theoretical and Applied Econometrics}, Chapter~4, pp.\  95--125. Springer.

\bibitem[\protect\citeauthoryear{Chen and Wang}{Chen and Wang}{2022}]{rrpack}
Chen, K. and W.~Wang (2022).
\newblock {\em rrpack: Reduced-Rank Regression}.
\newblock R package version 0.1-13.

\bibitem[\protect\citeauthoryear{Christiano, Eichenbaum, and Evans}{Christiano et~al.}{2005}]{christiano2005nominal}
Christiano, L.~J., M.~Eichenbaum, and C.~L. Evans (2005).
\newblock {Nominal rigidities and the dynamic effects of a shock to monetary policy}.
\newblock {\em Journal of Political Economy\/}~{\em 113}, 1--45.

\bibitem[\protect\citeauthoryear{Chudik and Pesaran}{Chudik and Pesaran}{2016}]{ChudikPesaran16}
Chudik, A. and M.~H. Pesaran (2016).
\newblock Theory and practice of {GVAR} modelling.
\newblock {\em Journal of Economic Surveys\/}~{\em 30}, 165--197.

\bibitem[\protect\citeauthoryear{Davidson}{Davidson}{2002}]{Davidson02}
Davidson, J. (2002).
\newblock {\em Stochastic Limit Theory\/} (2nd ed.).
\newblock Oxford: Oxford University Press.

\bibitem[\protect\citeauthoryear{Forni and Gambetti}{Forni and Gambetti}{2014}]{forni2014sufficient}
Forni, M. and L.~Gambetti (2014).
\newblock Sufficient information in structural {VAR}s.
\newblock {\em Journal of Monetary Economics\/}~{\em 66}, 124--136.

\bibitem[\protect\citeauthoryear{Forni, Giannone, Lippi, and Reichlin}{Forni et~al.}{2009}]{forni2009opening}
Forni, M., D.~Giannone, M.~Lippi, and L.~Reichlin (2009).
\newblock Opening the black box: {S}tructural factor models with large cross sections.
\newblock {\em Econometric Theory\/}~{\em 25}, 1319--1347.

\bibitem[\protect\citeauthoryear{Gold, Lederer, and Tao}{Gold et~al.}{2020}]{gold2020inference}
Gold, D., J.~Lederer, and J.~Tao (2020).
\newblock Inference for high-dimensional instrumental variables regression.
\newblock {\em Journal of Econometrics\/}~{\em 217\/}(1), 79--111.

\bibitem[\protect\citeauthoryear{Javanmard and Montanari}{Javanmard and Montanari}{2014}]{JavanmardMontanari14}
Javanmard, A. and A.~Montanari (2014).
\newblock Confidence intervals and hypothesis testing for high-dimensional regression.
\newblock {\em Journal of Machine Learning Research\/}~{\em 15}, 2869--2909.

\bibitem[\protect\citeauthoryear{Jord{\`a}}{Jord{\`a}}{2005}]{jorda2005estimation}
Jord{\`a}, {\`O}. (2005).
\newblock {Estimation and inference of impulse responses by local projections}.
\newblock {\em American Economic Review\/}~{\em 95\/}(1), 161--182.

\bibitem[\protect\citeauthoryear{Karapanagioti}{Karapanagioti}{2021}]{Karapanagioti21}
Karapanagioti, C. (2021).
\newblock Model selection for local projections instrumental variable methods - empirical application to government spending multipliers.
\newblock Master's thesis, Tilburg University, Tilburg, The Netherlands.

\bibitem[\protect\citeauthoryear{Kiefer and Vogelsang}{Kiefer and Vogelsang}{2005}]{kiefer_vogelsang_2005}
Kiefer, N.~M. and T.~J. Vogelsang (2005).
\newblock A new asymptotic theory for heteroskedasticity-autocorrelation robust tests.
\newblock {\em Econometric Theory\/}~{\em 21\/}(6), 1130–1164.

\bibitem[\protect\citeauthoryear{Kilian and L{\"u}tkepohl}{Kilian and L{\"u}tkepohl}{2017}]{kilian2017structural}
Kilian, L. and H.~L{\"u}tkepohl (2017).
\newblock {\em {Structural Vector Autoregressive Analysis}}.
\newblock Themes in Modern Econometrics. Cambridge University Press.

\bibitem[\protect\citeauthoryear{Kock and Callot}{Kock and Callot}{2015}]{KockCallot2015}
Kock, A.~B. and L.~Callot (2015).
\newblock Oracle inequalities for high dimensional vector autoregressions.
\newblock {\em Journal of Econometrics\/}~{\em 186}, 325--344.

\bibitem[\protect\citeauthoryear{Kock, Medeiros, and G}{Kock et~al.}{2020}]{kock2020penalized}
Kock, A.~B., M.~Medeiros, and V.~G (2020).
\newblock Penalized regressions.
\newblock In P.~Fuleky (Ed.), {\em Macroeconomic Forecasting in the Era of Big Data}, Volume~52 of {\em Advanced Studies in Theoretical and Applied Econometrics}, Chapter~7, pp.\  193--228. Springer.

\bibitem[\protect\citeauthoryear{Koop, Pesaran, and Potter}{Koop et~al.}{1996}]{koop1996impulse}
Koop, G., M.~H. Pesaran, and S.~M. Potter (1996).
\newblock Impulse response analysis in nonlinear multivariate models.
\newblock {\em Journal of Econometrics\/}~{\em 74}, 119--147.

\bibitem[\protect\citeauthoryear{Krampe, Paparoditis, and Trenkler}{Krampe et~al.}{2022}]{krampe2022}
Krampe, J., E.~Paparoditis, and C.~Trenkler (2022).
\newblock Structural inference in sparse high-dimensional vector autoregressions.
\newblock {\em Journal of Econometrics\/}, Forthcoming.

\bibitem[\protect\citeauthoryear{Lazarus, Lewis, Stock, and Watson}{Lazarus et~al.}{2018}]{lazarus2018har}
Lazarus, E., D.~J. Lewis, J.~H. Stock, and M.~W. Watson (2018).
\newblock Har inference: Recommendations for practice.
\newblock {\em Journal of Business \& Economic Statistics\/}~{\em 36\/}(4), 541--559.

\bibitem[\protect\citeauthoryear{Leeb and P{\"o}tscher}{Leeb and P{\"o}tscher}{2005}]{LeebPoetscher05}
Leeb, H. and B.~M. P{\"o}tscher (2005).
\newblock Model selection and inference: Facts and fiction.
\newblock {\em Econometric Theory\/}~{\em 21}, 21--59.

\bibitem[\protect\citeauthoryear{Li, Plagborg-Møller, and Wolf}{Li et~al.}{2021}]{li2021local}
Li, D., M.~Plagborg-Møller, and C.~K. Wolf (2021).
\newblock Local projections vs. vars: Lessons from thousands of dgps.
\newblock arXiv e-print 2104.00655.

\bibitem[\protect\citeauthoryear{Masini, Medeiros, and Mendes}{Masini et~al.}{2022}]{masini2019regularized}
Masini, R.~P., M.~C. Medeiros, and E.~F. Mendes (2022).
\newblock Regularized estimation of high-dimensional vector autoregressions with weakly dependent innovations.
\newblock {\em Journal of Time Series Analysis\/}~{\em 43}, 532--557.

\bibitem[\protect\citeauthoryear{McCracken and Ng}{McCracken and Ng}{2016a}]{mccracken2016fred}
McCracken, M.~W. and S.~Ng (2016a).
\newblock {FRED-MD}: a monthly database for macroeconomic research.
\newblock {\em Journal of Business \& Economic Statistics\/}~{\em 34}, 574--589.

\bibitem[\protect\citeauthoryear{McCracken and Ng}{McCracken and Ng}{2016b}]{McCrackenNg16}
McCracken, M.~W. and S.~Ng (2016b).
\newblock {FRED-MD}: A monthly database for macroeconomic research.
\newblock {\em Journal of Business \& Economic Statistics\/}~{\em 34}, 574--589.

\bibitem[\protect\citeauthoryear{Montiel~Olea and Plagborg-M{\o}ller}{Montiel~Olea and Plagborg-M{\o}ller}{2021}]{olea2020local}
Montiel~Olea, J.~L. and M.~Plagborg-M{\o}ller (2021).
\newblock Local projection inference is simpler and more robust than you think.
\newblock {\em Econometrica\/}~{\em 89}, 1789--1823.

\bibitem[\protect\citeauthoryear{Plagborg-Møller and Wolf}{Plagborg-Møller and Wolf}{2021}]{plagborg2021local}
Plagborg-Møller, M. and C.~K. Wolf (2021).
\newblock Local projections and {VARs} estimate the same impulse responses.
\newblock {\em Econometrica\/}~{\em 89}, 955--980.

\bibitem[\protect\citeauthoryear{Ramey}{Ramey}{2016}]{ramey2016macroeconomic}
Ramey, V.~A. (2016).
\newblock {Macroeconomic shocks and their propagation}.
\newblock In J.~B. Taylor and H.~Uhlig (Eds.), {\em Handbook of Macroeconomics}, Volume~2, pp.\  71--162. Elsevier.

\bibitem[\protect\citeauthoryear{Ramey and Zubairy}{Ramey and Zubairy}{2018}]{ramey2018government}
Ramey, V.~A. and S.~Zubairy (2018).
\newblock Government spending multipliers in good times and in bad: evidence from {US} historical data.
\newblock {\em Journal of Political Economy\/}~{\em 126}, 850--901.

\bibitem[\protect\citeauthoryear{Romer and Romer}{Romer and Romer}{2004}]{romer2004new}
Romer, C.~D. and D.~H. Romer (2004).
\newblock {A new measure of monetary shocks: derivation and implications}.
\newblock {\em American Economic Review\/}~{\em 94\/}(4), 1055--1084.

\bibitem[\protect\citeauthoryear{Sims}{Sims}{1980}]{sims1980macroeconomics}
Sims, C.~A. (1980).
\newblock Macroeconomics and reality.
\newblock {\em Econometrica\/}~{\em 48}, 1--48.

\bibitem[\protect\citeauthoryear{Sims and Zha}{Sims and Zha}{2006}]{sims2006were}
Sims, C.~A. and T.~Zha (2006).
\newblock Were there regime switches in {U.S.} monetary policy?
\newblock {\em American Economic Review\/}~{\em 96\/}(1), 54--81.

\bibitem[\protect\citeauthoryear{Stock and Watson}{Stock and Watson}{2016}]{stock2016DFM}
Stock, J. and M.~Watson (2016).
\newblock Dynamic factor models, factor-augmented vector autoregressions, and structural vector autoregressions in macroeconomics.
\newblock In J.~B. Taylor and H.~Uhlig (Eds.), {\em Handbook of Macroeconomics}, Volume~2, pp.\  415--525. Elsevier.

\bibitem[\protect\citeauthoryear{Stock and Watson}{Stock and Watson}{2018}]{stock2018identification}
Stock, J.~H. and M.~W. Watson (2018).
\newblock {Identification and estimation of dynamic causal effects in macroeconomics using external instruments}.
\newblock {\em Economic Journal\/}~{\em 128}, 917--948.

\bibitem[\protect\citeauthoryear{Uematsu and Yamagata}{Uematsu and Yamagata}{2022a}]{uematsu2022estimation}
Uematsu, Y. and T.~Yamagata (2022a).
\newblock Estimation of sparsity-induced weak factor models.
\newblock {\em Journal of Business \& Economic Statistics\/}, forthcoming.

\bibitem[\protect\citeauthoryear{Uematsu and Yamagata}{Uematsu and Yamagata}{2022b}]{uematsu2022inference}
Uematsu, Y. and T.~Yamagata (2022b).
\newblock Inference in sparsity-induced weak factor models.
\newblock {\em Journal of Business \& Economic Statistics\/}, forthcoming.

\bibitem[\protect\citeauthoryear{van~de Geer, B{\"u}hlmann, Ritov, and Dezeure}{van~de Geer et~al.}{2014}]{vandeGeer14}
van~de Geer, S., P.~B{\"u}hlmann, Y.~Ritov, and R.~Dezeure (2014).
\newblock On asymptotically optimal confidence regions and tests for high-dimensional models.
\newblock {\em Annals of Statistics\/}~{\em 42}, 1166--1202.

\bibitem[\protect\citeauthoryear{van~de Geer}{van~de Geer}{2016}]{vandeGeer2016book}
van~de Geer, S.~A. (2016).
\newblock {\em Estimation and testing under sparsity}.
\newblock Springer.

\bibitem[\protect\citeauthoryear{Wilms, Matteson, Bien, Basu, Nicholson, and Wegner}{Wilms et~al.}{2021}]{bigtime}
Wilms, I., D.~S. Matteson, J.~Bien, S.~Basu, W.~Nicholson, and E.~Wegner (2021).
\newblock {\em bigtime: Sparse Estimation of Large Time Series Models}.
\newblock R package version 0.2.1.

\bibitem[\protect\citeauthoryear{Yamada}{Yamada}{2017}]{yamada2017}
Yamada, H. (2017).
\newblock {The Frisch–Waugh–Lovell theorem for the lasso and the ridge regression}.
\newblock {\em Communications in Statistics-Theory and Methods\/}~{\em 46}, 10897--10902.

\end{thebibliography}

\section*{Appendix A: Assumptions}
\renewcommand{\theequation}{A.\arabic{equation}}
\renewcommand{\theassumption}{A.\arabic{assumption}}
\def\thesection{\Alph{section}}
\setcounter{section}{0}

In the appendix, we use the following additional notation.
For a matrix $\bA$, we let $\norm{\bA}_r = \max_{\norm{\bx} = 1} \norm{\bA \bx}_r$ for any $r \in [0, \infty]$ and $\norm{\bA}_{\max}=\max_{i,j}\left\vert a_{i,j}\right\vert$. $\Lambda_{\min}(\bA)$ denotes the minimum eigenvalue of $\bA$.
We 
frequently make use of arbitrary positive finite constants $C$ (or its sub-indexed version $C_i$) whose values may change from line to line throughout the paper, but they are always independent of the time and cross-sectional dimension.
We say a sequence $\eta_T$ is of size $-x$ if $\eta_T=O\left(T^{-x-\varepsilon}\right)$ for some $\varepsilon>0$. 

\begin{assumption}\label{ass:DGP}
There exist some constants $\bar m>m>2$, and $d\geq \max\{1,(\bar m/m-1)/(\bar m-2)\}$ such that
\begin{enumerate}[label=(\alph*)]
\item\label{ass:DGPStationary} $(\boldsymbol{x}_t^\prime, u_t)^\prime$ is a mean zero  process with $\E \bx_t u_t=\boldsymbol{0}$ and\\ $\max_{1\leq j\leq N}\E\abs{x_{j,t}}^{2\bar m} \leq C$, $\E\abs{u_t}^{2\bar m} \leq C$, 
\item\label{ass:DGPvMoments} $\max_{1\leq j\leq S}\E\abs{v_{j,t}}^{2\bar m} \leq C$,
\item\label{ass:DGPNED}Let $\{\bepsilon_{T,t}\}$ denote a $k(T)$-dimensional triangular array that is $\alpha$-mixing of size $-d/(1/m-1/\bar{m})$ with $\sigma\text{-field}$ $\mathcal{F}^{\bepsilon}_t:=\sigma\left\lbrace\bepsilon_{T,t},\bepsilon_{T,t-1},\dots\right\rbrace$ such that $(\boldsymbol{x}_t^\prime, u_t)^\prime$ is $\mathcal{F}^{\bepsilon}_t$-measurable. The processes $u_t$ and $x_{j,t}$ are $L_{2m}$-near-epoch-dependent (NED) of size $-d$ on $\bepsilon_{T,t}$ with positive bounded constants, uniformly over $j=1,\ldots,N$.
\end{enumerate}
\end{assumption}

\begin{assumption}
\label{ass:Sparsity}
For some $0\leq r<1$ and sparsity level $s_r$, define the $N$-dimensional sparse compact parameter space
$\bB_{N}(r, s_r) :=\left\lbrace \bbeta\in \mathds{R}^{N}: \norm{\bbeta}_r^r \leq s_r, \; \norm{\bbeta}_{\max} \leq C, \, \exists C < \infty \right\rbrace$.
Then (a) $\boldsymbol{\beta}\in\boldsymbol{B}_{N}(r, s_r)$ and (b) $\bgamma_j\in\bB_{N-1}(r,s_{r,j})$ for all $j\in \cH$, where $\bgamma_j=\argmin_{\bgamma_j^*\in\mathds{R}^{N-1}}\E\norm{\bx_j-\bX_{-j}\bgamma_j^*}_2^2/T$.
\end{assumption}

\begin{assumption}\label{ass:Eigenvalue}
Let $\Lambda_{\min}(\bSigma)$ denote the smallest eigenvalue of $\bSigma=\E\bX^\prime\bX/T$, and $\Lambda_{\min}(\boldsymbol{\Omega}_{N,T})$ the smallest eigenvalue of 
${\boldsymbol{\Omega}}_{T} := \E\left[\frac{1}{T}\left(\sum_{t=1}^T \bq_t\right)\left(\sum_{t=1}^T \bq'_t\right)\right]$ Assume that $1/C\leq\Lambda_{\min}(\bSigma)\leq C$, and $1/C\leq\Lambda_{\min}(\boldsymbol{\Omega}_{N,T})\leq C$.
\end{assumption}

\begin{assumption}\label{ass:HFinite}
$H\leq C$, where $H$ is the cardinality of $\cH$. 
\end{assumption}

\begin{assumption}\label{ass:asymptoticrates} Let
$\lambda_{\min}:=\min\{\lambda,\min_{j}\lambda_j\}$, $\lambda_{\max}:=\max\{\lambda,\max_{j}\lambda_j\}$, and $s_{r,\max}:=\max\{s_r,\max_{j}s_{r,j}\}$. Then  $Q_TT^{-\frac{1}{2/d+2m/(m-2)}}\to0$ for $Q_T\to\infty$, $\lambda\sim\lambda_{\max}\sim\lambda_{\min}$, 
and
\begin{equation*}
\begin{split}
0<r<1:\quad &(\ln \ln T)^{-1} s_{r,\max}^{1/r}\left[\frac{N^{\left(\frac{2}{d}+\frac{2}{m-1}\right)}}{\sqrt{T}}\right]^{\frac{1}{r\left(\frac{1}{d}+\frac{m}{m-1}\right)}}\leq\lambda\\
&\quad\leq\ \ln \ln T \left[Q_T^2\sqrt{T}s_{r,\max}\right]^{-1/(2-r)},\\
r=0:\quad &(\ln \ln T)^{-1} \frac{N^{1/m}}{\sqrt{T}}\leq\lambda\leq \ln \ln T \left[Q_T^2\sqrt{T}s_{0,\max}\right]^{-1/2}.
\end{split}
\end{equation*}
These bounds are feasible when $Q_T^rs_{r,\max}N^{\left(2-r\right)\left(\frac{d+m-1}{dm+m-1}\right)}T^{\frac{1}{4}\left(r-\frac{d(m-1)(2-r)}{dm+m-1}\right)}\to0$, and additionally when $Q_T^2s_{0,\max}\frac{N^{2/m}}{\sqrt{T}}\to0$ if $r=0$.
\end{assumption}

\section*{Appendix B: Proofs} \label{app:proofs}
\renewcommand{\theequation}{B.\arabic{equation}}
\renewcommand{\thelemma}{B.\arabic{lemma}}
\def\thesection{\Alph{section}}
\setcounter{section}{0}

\begin{lemma}\label{lma:compatibility} Take a vector $\left\lbrace \bz\in\mathds{R}^{N}:\norm{\bz_{\cH^c}}_1\leq 3\norm{\bz_{\cH}}_1 \right\rbrace$, an index set $\cH$ with cardinality $\abs{\cH}$ and define the subscript operator $\bz_\cH:=\{\bz^*\in\mathds{R}^N:z^*_{j}=z_{j}\mathds{1}_{\{j\in \cH\}}\}$. Under \cref{ass:Eigenvalue}, on the set $\setCC(\cH):=\left\lbrace\norm{\hat\bSigma-\bSigma}_{\max}\leq C/\abs{\cH}\right\rbrace$,
\begin{equation*}
\Vert \bz_{\cH}\Vert_1\leq\sqrt{\frac{2\vert \cH\vert \bz'\hat{{\boldsymbol{\Sigma}}}\bz}{\Lambda_{\min}}}
\end{equation*}
\end{lemma}

\medskip \noindent
\textbf{Proof:}
By Lemma 6.17 and Corollary 6.8 in \citet{SHD11}, this result holds for the ``$\bSigma$-compatibility condition''. By Lemma 6.23 in \citet{SHD11}, this compatibility condition holds under \cref{ass:Eigenvalue}, and the result follows from the fact that $\Lambda_{\min}$ bounds the compatibility constant from below.\hfill$\square$

\medskip
\begin{lemma}\label{lma:GeneralOracle}
Let $P$ be an index set with cardinality $\abs{P}$ such that $\cH\subseteq P$.
Define the set $\setEP{T}(z)=\left\lbrace\max_{j\leq N,s\leq T}\left[\left\vert\sum_{t=1}^{s}u_t x_{j,t}\right\vert\right]\leq z\right\rbrace$. 
Under \cref{ass:Eigenvalue}, on 
$\setEP{T}(T\frac{\lambda}{4})\bigcap\setCC(P)$:
\begin{equation*}\begin{split}
T^{-1}\norm{\boldsymbol{X}(\hat{\boldsymbol{\beta}}^{(L)}-{\boldsymbol{\beta}})}_2^2+\lambda/4 \norm{\hat{\boldsymbol{\beta}}^{(L)}-{\boldsymbol{\beta}}}_1\leq&C_1\lambda^2\vert P\vert+C_2\lambda\norm{{\boldsymbol{\beta}}_{P^c}}.\\
\end{split}\end{equation*}
\end{lemma}

\medskip \noindent
\textbf{Proof:}
The proof closely follows the proof of Lemma A.6 in \citet{adamek2021lasso}, 
based on Theorem 2.2 of \citet{vandeGeer2016book}, with small modifications to handle the penalty matrix $\Weight$. 
From the Lasso optimization problem in \cref{eq:problem}, we have the Karush-Kuhn-Tucker (KKT) conditions
$T^{-1}\boldsymbol{X}'({\boldsymbol{y}}-\boldsymbol{X}\hat{\boldsymbol{\beta}}^{(L)})=\bW\hat\kappa^*\lambda,$
where $\hat\kappa^*=(\bzero_{1\times H},\hat\kappa^\prime)^\prime$, and $\hat\kappa$ is the subdifferential of $\Vert\hat{\boldsymbol{\beta}}^{(L)}_{-\cH}\Vert_1$. This then leads to the inequality 
\begin{equation*}\begin{split}
T^{-1}({\boldsymbol{\beta}}-\hat{\boldsymbol{\beta}}^{(L)})'\boldsymbol{X}'({\boldsymbol{y}}-\boldsymbol{X}\hat{\boldsymbol{\beta}}^{(L)})
&= ({\boldsymbol{\beta}}-\hat{\boldsymbol{\beta}}^{(L)})^\prime\bW\hat\kappa^*\lambda
={{\boldsymbol{\beta}}}^{\prime}\bW\hat\kappa^*\lambda-\hat\bbeta^{(L)\prime}\bW\hat\kappa^*\lambda\\
&\leq\lambda\Vert\bW^\prime\bbeta \Vert_1-\lambda\Vert\bW^\prime\hat\bbeta^{(L)}\Vert_1.
\end{split}\end{equation*}
To deal with the matrix $\Weight$ in these expressions, we use the bound  $\norm{\bW^\prime(\hat\bbeta_P^{(L)}-\bbeta_P)}_1\leq\norm{\hat\bbeta_P^{(L)}-\bbeta_P}_1$ and the property that $\norm{\Weight^\prime\bz_{P^c}}_1=\norm{\bz_{P^c}}_1$ for $\bz\in\mathds{R}^{N}$, since $\cH\subseteq P$, and those elements set to zero by $\Weight$ are already zero due to subscript $P^c$ operator. Using these additional arguments, we can show that
\begin{equation*}
T^{-1}\norm{\bX(\hat\bbeta^{(L)}-\bbeta)}_2^2 
\leq\frac{5\lambda}{4}\norm{ \hat\bbeta_P^{(L)}-\bbeta_P}_1-\frac{3\lambda}{4}\norm{ \hat\bbeta_{P^c}^{(L)}-\bbeta_{P^c}}_1+2\lambda\norm{\bbeta_{P^c}}_1,
\end{equation*}
and the remainder of the proof is analogous to the proof of Lemma A.6 in \citet{adamek2021lasso},
using \cref{lma:compatibility} instead of Lemma A.5 in \citet{adamek2021lasso}. \hfill$\square$

\medskip
\begin{lemma}\label{thm:ErrorBound}
Under \cref{ass:DGP,ass:Sparsity,ass:Eigenvalue,ass:HFinite}, for any
\begin{equation}
\begin{split}
0<r<1:\quad &\lambda\geq  C\ln(\ln( T))^{\frac{d+m-1}{r(dm+m-1)}}\left[s_r\left(\frac{N^{\left(\frac{2}{d}+\frac{2}{m-1}\right)}}{\sqrt{T}}\right)^{\frac{1}{\left(\frac{1}{d}+\frac{m}{m-1}\right)}}\right]^{\frac{1}{r}}\\
r=0:\quad &s_0\leq C \ln(\ln( T))^{-\frac{d+m-1}{dm+m-1}}\left[\frac{\sqrt{T}}{N^{\left(\frac{2}{d}+\frac{2}{m-1}\right)}}\right]^{\frac{1}{\left(\frac{1}{d}+\frac{m}{m-1}\right)}},\\
&\lambda\geq C{ \ln(\ln( T))}^{1/m}\frac{N^{1/m}}{\sqrt{T}}
\end{split}
\end{equation}
and when $N,T$ are sufficiently large, the following holds with probability at least $1-\probseqCC-\probseqE$:
\begin{equation*}
\text{(i)} \quad \frac{1}{T} \norm{\boldsymbol{X}(\hat{\boldsymbol{\beta}}^{(L)}-{\boldsymbol{\beta}})}_{2}^2
\leq C\lambda^{2-r}s_r, \qquad 
\text{(ii)} \quad\norm{\hat{\boldsymbol{\beta}}^{(L)}-{\boldsymbol{\beta}}}_1
\leq C\lambda^{1-r}s_r,
\end{equation*}
where $\probseqCC$ and $\probseqE$ are sequences that converge to 0 for large $T,N$ (defined formally in the proof).
\end{lemma}

\medskip \noindent
\textbf{Proof:}
Let $P_\lambda:=\{j:\abs{\beta_j}>\lambda\}$, and $P_{\cH,\lambda}:= \cH\bigcup P_\lambda$,
such that by construction, $\cH\subseteq P_{\cH,\lambda}$. It then follows that by \Cref{ass:Eigenvalue} and \cref{lma:GeneralOracle}, 
we have on the set	$\setEP{T}(T\frac{\lambda}{4})\bigcap\setCC(P_{\cH,\lambda})$
\begin{equation*}\begin{split}
\norm{\bX(\hat\bbeta^{(L)}-\bbeta)}_2^2/T+\frac{\lambda}{4}\Vert\hat{\boldsymbol{\beta}}^{(L)}-{\boldsymbol{\beta}}\Vert_1\leq&C_1\lambda^2\vert P_{\cH,\lambda}\vert+C_2\lambda\Vert{\boldsymbol{\beta}}_{P_{\cH,\lambda}^c}\Vert_1.\\
\end{split}\end{equation*}
It follows directly from \cref{ass:Sparsity} that 
\begin{equation*}\begin{split}
\vert P_{\cH,\lambda}\vert&
\leq H+\sum_{j=1}^{N}\mathds{1}_{\{\abs{\beta_j}>\lambda\}}\left(\frac{\abs{ \beta_j}}{\lambda}\right)^r
\leq H+\lambda^{-r}\sum_{j=1}^{N} \abs{\beta_l}^r =H+\lambda^{-r}s_r.\\
\end{split}\end{equation*}
Note that since $H\leq C$ by \cref{ass:HFinite} and both $\lambda^{-r}$ and $s_r$ asymptotically grow as $N,T\to \infty$, for sufficiently large $N$ and $T$, $H+\lambda^{-r}s_r\leq C\lambda^{-r}s_r$. Following the proof of Lemma A.7 in \citet{adamek2021lasso}, 
we can also show that $\norm{\boldsymbol{\beta}_{P_{\cH,\lambda}^c}}_1
\leq\lambda^{1-r}s_r$, and obtain the error bound
\begin{align*}
T^{-1} \Vert \boldsymbol{X}(\hat{\boldsymbol{\beta}}^{(L)}-{\boldsymbol{\beta}})\Vert_2^2 +\lambda\Vert\hat{\boldsymbol{\beta}}^{(L)}-{\boldsymbol{\beta}}\Vert_1 &\leq C_1\lambda^{2-r}s_r+C_2\lambda^{2-r}s_r
=C\lambda^{2-r}s_{r}.
\end{align*}
For the set $\setCC(P_{\cH,\lambda})$, under \cref{ass:DGP,ass:Sparsity,ass:Eigenvalue}, we can apply Lemma A.3 in \citet{adamek2021lasso} 
with $\eta_T=1/\ln(\ln(T))^{\frac{dm+m-1}{d+m-1}}$ to show that $\P\left(\setCC(P_\lambda)\right)\geq 1-3\left[1/\ln(\ln(T))\right]^{\frac{dm+m-1}{d+m-1}}:=1-\probseqCC\to1$ as $N,T\to\infty$. 
Note that the condition $\eta_T\leq\frac{N^2}{e}$ is satisfied for sufficiently large $N,T$, and we also use this in the proof of Lemma A.3 in \citet{adamek2021lasso} 
to plug in $\vert P_{\cH,\lambda}\vert\leq H+\lambda^{-r}s_r\leq C\lambda^{-r}s_r$.

Regarding the set $\setEP{T}(T\frac{\lambda}{4})$, by \cref{ass:DGP} and Lemma A.4 in \citet{adamek2021lasso}, 
$\P\left(\setEP{T}(T\lambda/4)\right)\geq 1-CN(\sqrt{T}\lambda)^{-m}:=1-\probseqE$
By the union bound, 
\begin{equation*}
    \P\left(\setEP{T}(T\lambda/4)\cap\setCC(P_{\cH,\lambda})\right)\geq 1-\probseqE-\probseqCC,
\end{equation*}
and therefore the error bound holds with this probability as well. With the error bound, items (i) and (ii) follow straightforwardly. \hfill $\square$

\medskip \noindent
\textbf{Proof of \cref{thm:Inference}:}
This result follows from Corollary 2 of \citet{adamek2021lasso}, 
with some small differences summarized here. First, in \citet{adamek2021lasso}, 
the tested hypotheses involve the full vector of parameters $\bbeta$, whereas our result only holds for $\bbeta_{\cH}$. However, the matrix $\bR_N$ in \citet{adamek2021lasso} 
is restricted to have only $H$ nonzero columns such that $\left\lbrace j:\sum_{p=1}^P\abs{r_{p,j}}>0\right\rbrace=\cH$. The setup here is therefore equivalent to taking $\bR_N=\left(\underset{H\times H}{\bI},\underset{H\times(N-H)}{\bzero}\right)$, and the elements we omit here are equal to zero. 

Second, parts of the proof which require bounds on $\norm{\hat\bbeta^{(L)}-\bbeta}_1$ or $\norm{\hat\bu^{(L)}-\bu}_{2}$ need to be addressed by the new lasso error bound in \cref{thm:ErrorBound}, rather than by 
Corollary 1 in \citet{adamek2021lasso}. 
Specifically, in Lemma B.8 of \citet{adamek2021lasso}
we need the bound $\norm{\hat\bbeta^{(L)}-\bbeta}_{1}\leq C\lambda^{1-r}s_r$, and in
Lemma B.13 of \citet{adamek2021lasso}
we need $\norm{\hat{\boldsymbol{u}} - \boldsymbol{u}}_2 = \norm{\boldsymbol{X} \left(\hat{\boldsymbol{\beta}}^{(L)} - \boldsymbol{\beta} \right)}_2 \leq C \sqrt{T \lambda^{2-r} s_r}$. Note that since the new lasso error bound is the same as for the regular lasso, no changes to these proofs are necessary. \hfill $\square$
\normalsize
\clearpage

\renewcommand{\theequation}{C.\arabic{equation}}
\renewcommand{\thesection}{C}
\renewcommand{\thefigure}{C.\arabic{figure}}
\renewcommand{\thetable}{C.\arabic{table}}

\setcounter{equation}{0}
\setcounter{figure}{0}
\setcounter{table}{0}
\setcounter{section}{0}

\section*{Appendix C: Supplementary} \label{app:supplementary}

 \subsection{Simulations: Implementation Details and Extra Figures}\label{app:extra_figures}
 {This section contains the implementation details and additional figures for our simulation experiments, \Cref{sec:Simulations}. 
 }
 
 \subsubsection{Sparse Structural VAR Model}\label{app:sparse_var}
 {
Consider the DGP
\begin{equation}\label{eq:sim}
    \left(\begin{array}{c}
        y_t\\
        \bw_{s,t}  
    \end{array}\right)=\underset{ P\times 1}{\bz_t}=\sum\limits_{k=1}^4\bA_{k}\bz_{t-k}+\bepsilon_t,\quad \bepsilon_t\overset{iid}{\sim}N(\bzero,\bI),
\end{equation}
where the 
autoregressive parameter matrices $\bA_k$ are tapered Toeplitz matrices
for which we consider two different settings.
In the first setting, we take
$(\bA_k)_{i,j}=\rho_{k}^{\abs{i-j}+1}$ if $\abs{i-j}<P/2$ and 0 otherwise, and $(\rho_1,\rho_2,\rho_3,\rho_4)=(0.2,0.15,0.1,0.05)$. 
In the second setting, we simply switch the signs of all entries in the second and fourth autoregressive parameter matrices. We consider different values for the number of variables $P=\{20, 40, 100\}$ and time series length $T =\{100, 200, 500\}$.\footnote{The number of regressors in the model is $N=5 P$, due to the included lags.} 
}
\Cref{eq:sim} can be seen as
a simple structural VAR
with structural shocks $\bepsilon_t$.
The true impulse responses are obtained by inverting the lag polynomial, rewriting it as  $\bx_t=\sum_{k=0}^\infty\bB_{k}\bepsilon_{t-k}=\bB(L)\bepsilon_t$, where $\bB(L)=\bA(L)^{-1}=\left(\bI-\sum_{k=1}^4\bA_k L^k\right)^{-1}$ and $\bB_{0}=\bI$.  
We analyse the response of $y_t$ to the first shock $\epsilon_{1,t}$ using local projections
\begin{equation*}
    y_{t+h}=\phi_{h}y_{t}+ \boldsymbol{\eta}_h^\prime\bw_{s,t}+\sum_{k=1}^K\boldsymbol{\delta}_{h,k}^\prime\bz_{t-k}+u_{h,t}, \quad h=1,\dots,h_{\max}, 
\end{equation*}
and obtain estimates of the impulse responses $\hat\phi_{1}, 
\dots, \hat\phi_{h_{\max}}$ with $h_{\max}=10$ and $K=4$.\footnote{At horizon 0 the response is $(\bB_0)_{1,1}=1$ by our identification scheme, and thus does not need to be estimated.}

\begin{figure}
\centering
\includegraphics[width=0.95\textwidth]
{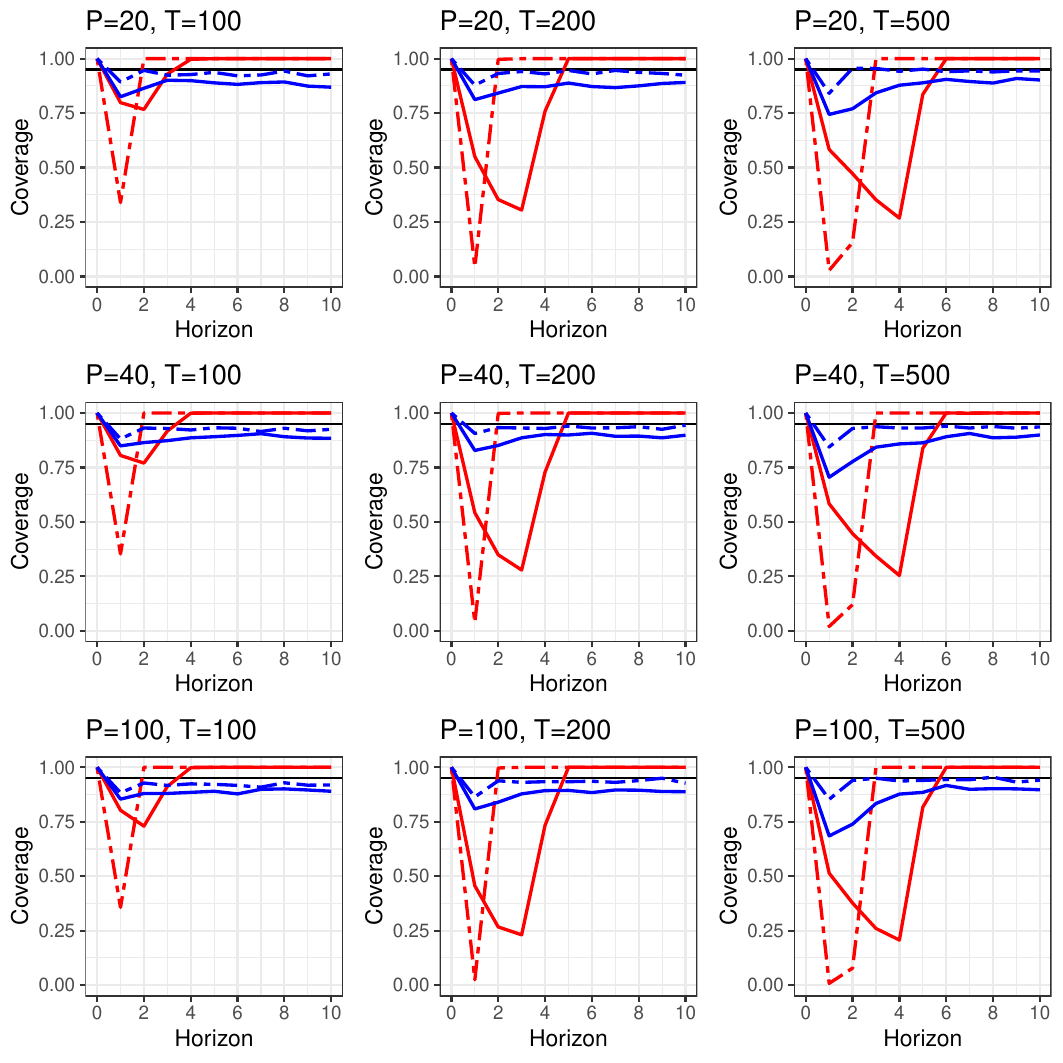}
\caption{Coverage of the standard desparsified lasso (red) and the proposed desparsified lasso with $\phi_{h}$ unpenalized (blue). Dashed lines indicate results for the sign-switching DGP.}\label{fig:sim_coverages}\label{fig:S1}
\end{figure}

\begin{figure}[!ht]
\centering
\includegraphics[width=0.95\textwidth]
{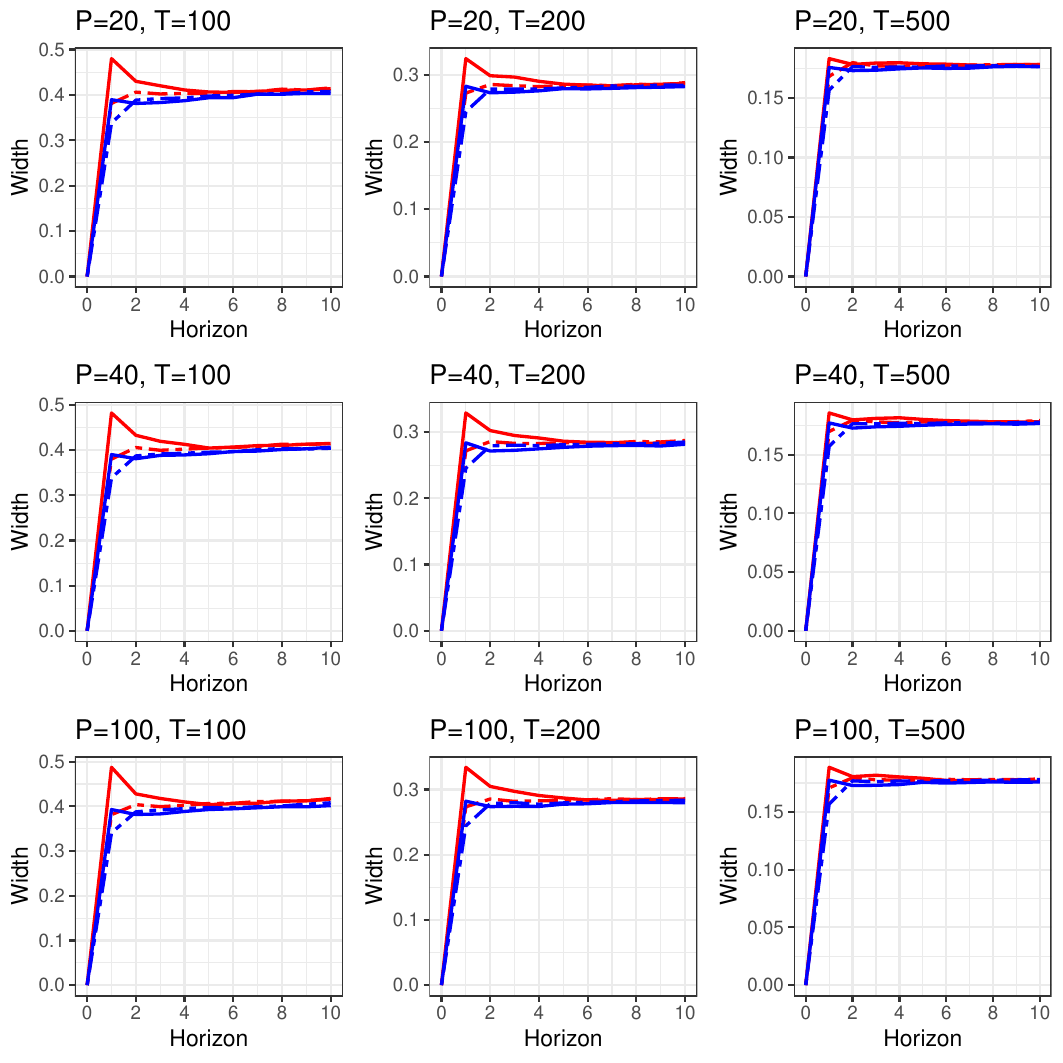}
\caption{Interval widths of the standard desparsified lasso (red) and the proposed desparsified lasso with $\phi_{h}$ unpenalized (blue). Dashed lines indicate results for the sign-switching DGP.}\label{fig:sim_widths}\label{fig:S2}
\end{figure}

\Cref{fig:sim_coverages} presents the coverage rates all values of $P$ and $T$ of the proposed desparsified lasso (blue), which leaves the impulse response parameters of interest unpenalized, in comparison to the standard desparsified lasso (red). 
\Cref{fig:sim_widths} displays the corresponding interval widths.

\subsubsection{Empirically Calibrated Dynamic Factor Model}\label{sec:DFM_details}
{
We consider an empirically calibrated dynamic factor model in the spirit of \cite{lazarus2018har} and \cite{li2021local}, which we estimate 
based on 122 monthly variables $\bx_t$ from the
FRED-MD database (\citealp{mccracken2016fred}).\footnote{The variables are detailed in \cref{sec:data}.}
}
The DFM is given by
\begin{equation}\label{eq:factors}
    \bx_t=\bLambda\bbf_t+\bv_t,
\end{equation}
\vspace{-1cm}
\begin{equation}\label{eq:VAR}
    \bbf_t=\bPhi\bbf_{t-1}+\bH\bepsilon_t,
\end{equation}
\vspace{-1cm}
\begin{equation}\label{eq:ARs}
   v_{i,t}= \Delta_i(L)v_{i,t-1}+\Xi_i \xi_{i,t},
\end{equation}
where \cref{eq:factors} describes the 6-factor model for $\bX_t$, \cref{eq:VAR} describes the VAR(1) model for the factors, and \cref{eq:ARs} gives the individual AR(2) models for the idiosyncratic errors from the factor model. Note that we can re-write \cref{eq:ARs} as $\bv_t=\bDelta(L)\bv_{t-1}+\bXi\bxi_t$, with $\bDelta$ and $\bXi$ diagonal.

\begin{figure}[!ht]
\centering
\includegraphics[width=0.95\textwidth]
{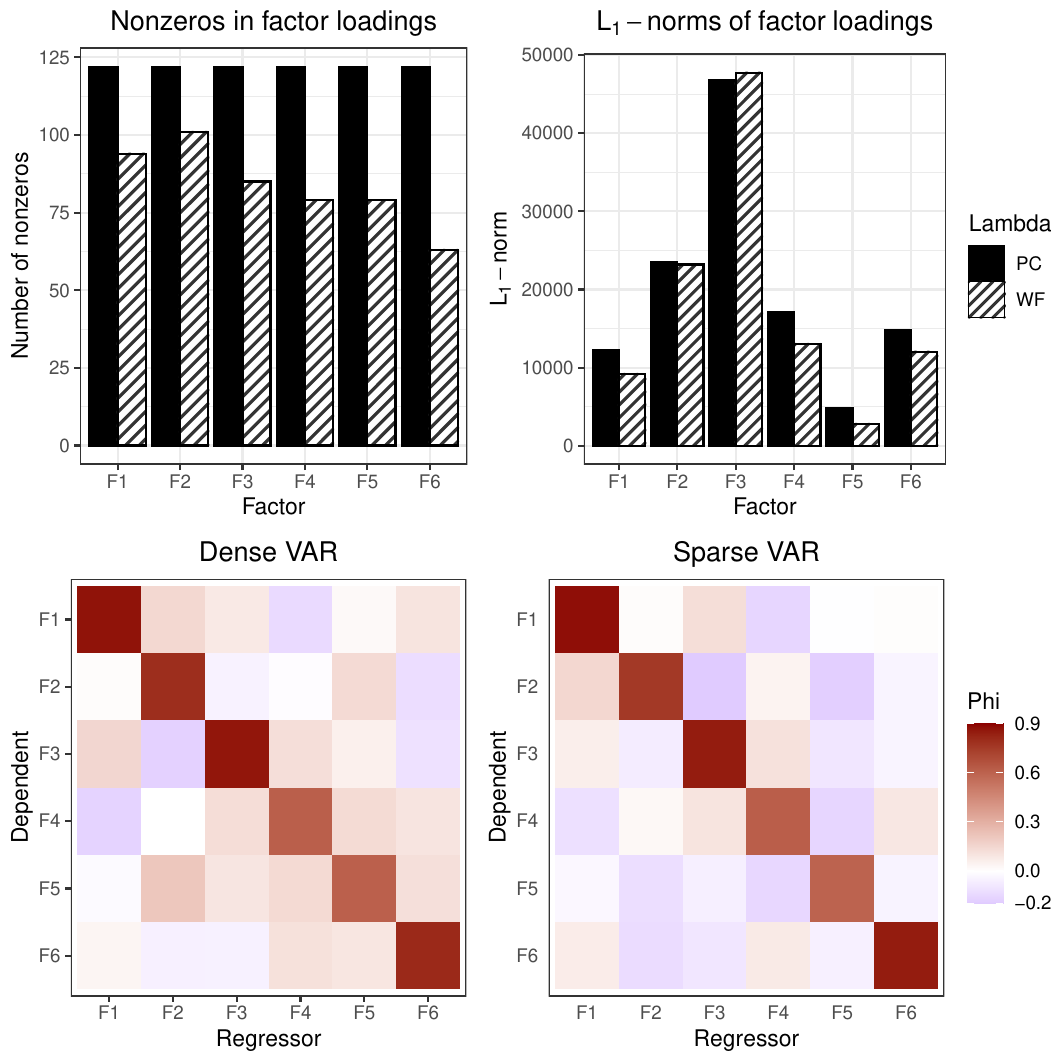}
\caption{Differences in sparsity between the dense and sparse DFM models.}\label{fig:dense_vs_sparse_DFM}\label{fig:S3}
\end{figure}

We consider two different versions of this DFM: A ``dense'' DFM where we set $\bLambda = \hat{\bLambda}_{PC}$ and $\bPhi=\hat\bPhi_{OLS}$, with with $\hat{\bLambda}_{PC}$ obtained using principal components and $\hat\bPhi_{OLS}$ obtained by fitting a VAR(1) on the estimated factors by OLS; and a ``sparse'' DFM, where $\bLambda = \hat{\bLambda}_{WF}$ and $\bPhi=\hat\bPhi_{L}$, with $\hat{\bLambda}_{WF}$ obtained using the WF-SOFAR estimator of \cite{uematsu2022estimation} and $\hat\bPhi_{L}$ by fitting the VAR(1) with the lasso.\footnote{We use the WF-SOFAR estimator with BIC criterion as implemented in the R package \texttt{rrpack} (\citealp{rrpack}), see Section D of the supplementary material to \cite{uematsu2022estimation} for further details. For the lasso VAR, we use the estimator with BIC criterion implemented in the R package \texttt{bigtime} (\citealp{bigtime}).
} To see the differences in sparsity between the dense and sparse specifications, see \Cref{fig:dense_vs_sparse_DFM}; the top row of plots compares the number of nonzero entries and $L_1$-norms for each column of $\bLambda$, the bottom shows heatmaps of $\bPhi$.

After fitting the models as described above, we use the residuals 
$\hat\be_t=\hat\bbf_t-\hat\bPhi \hat\bbf_{t-1}$ and $\hat\bzeta_t=\hat\bv_t-\hat\bDelta(L)\hat\bv_{t-1}$ to compute the sample covariances $\hat\bSigma_{\be}=\frac{1}{T-1}\sum_{t=2}^T \hat\be_t \hat\be_t'$ and $\hat \bSigma_{\bzeta}$ with $\hat\sigma_{\bzeta,i}^2=\frac{1}{T-2}\sum_{t=3}^T\hat\zeta_{i,t}^2$ on the diagonal and 0 otherwise. Following \cite{li2021local}, we then choose $\bH$ as a function of $\hat\bLambda$ and $\hat\bSigma_{\be}$ which maximises the contemporaneous effect of $\epsilon_{1,t}$ on the Federal Funds Rate (FFR), the monetary policy instrument in this simulation; for details, see Appendix A.2 of \cite{li2021local}. We take $\bXi=\hat\bSigma_{\bzeta}^{-1/2}$. We use a recursive identification scheme similar to that of \cite{li2021local}, where we order Industrial Production (IP) first, FFR last, and estimate the impulse response of IP to FFR. As noted in Section 3.2 of \cite{li2021local}, the shock identified in this way is not necessarily structural, i.e., it does not necessarily correspond to any specific shock in $\bepsilon_t$ or $\bxi_t$.
To compute the true impulse response implied by the calibrated model, we use the state-space representation described in Appendix A.4 of \cite{li2021local}. The data in the simulations are generated as in \cref{eq:factors,eq:VAR,eq:ARs} with $\bepsilon_t\overset{iid}{\sim}N(\bzero,\bI)$ and $\bxi_t\overset{iid}{\sim}N(\bzero,\bI)$, at sample lengths $T=\left\lbrace 200,400,600\right\rbrace$  with 1000 replications.

\begin{figure}[t]
\centering
\includegraphics[width=\textwidth]{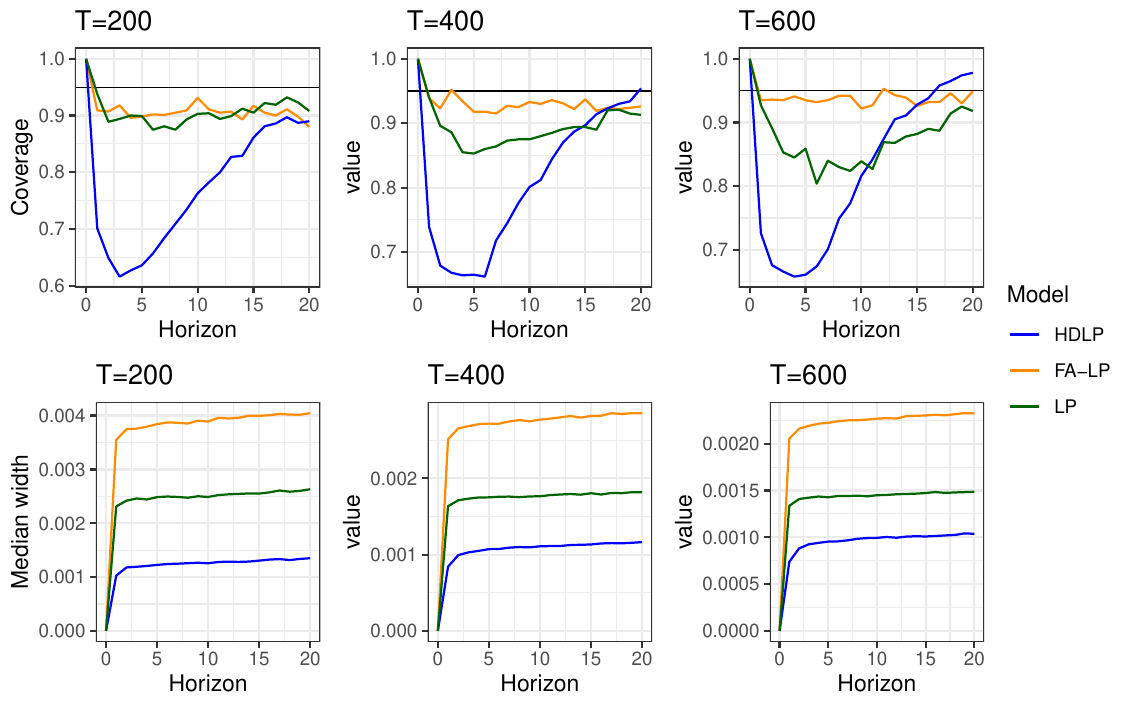}
\caption{Coverage rates and median interval widths
in the dense DFM.}\label{fig:dense_models}\label{fig:S4}
\end{figure}

\subsubsection{Sensitivity of HDLP to the Long-run Covariance Estimator}\label{sec:other_LRVs}

\begin{figure}[t]
\centering
\includegraphics[width=\textwidth]{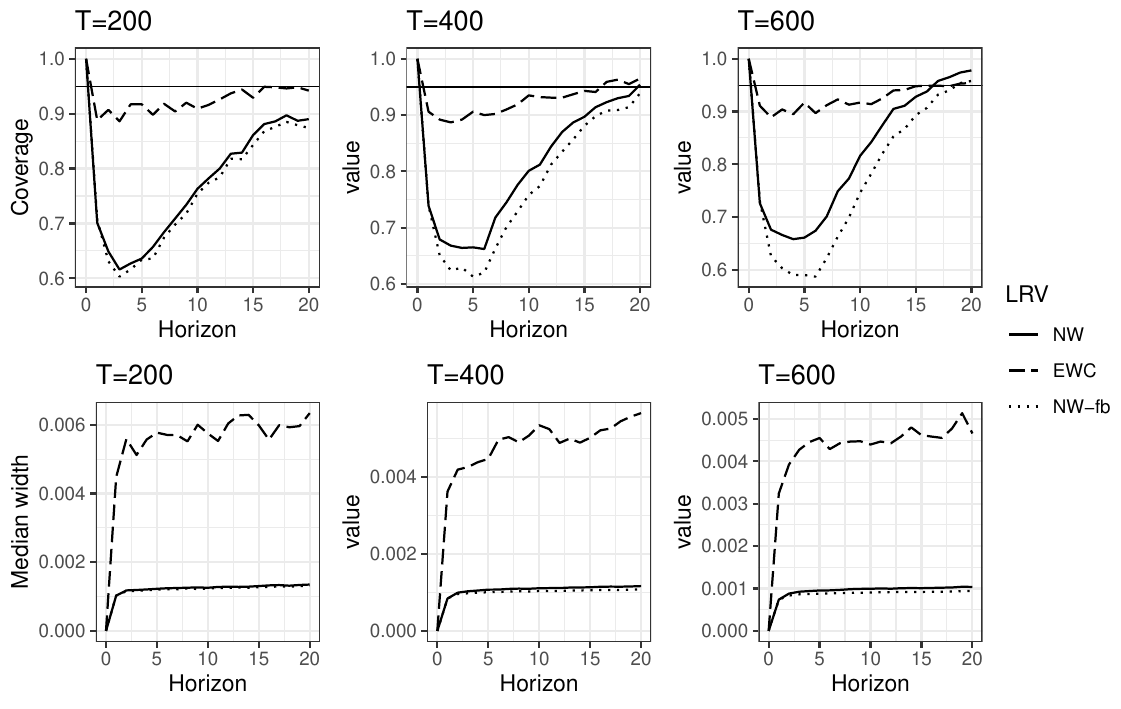}
\caption{Coverage rates and median interval widths of HDLP with different long-run covariance estimators in the dense DFM.}\label{fig:dense_lrvs}\label{fig:S5}
\end{figure}

\begin{figure}[t]
\centering
\includegraphics[width=\textwidth]{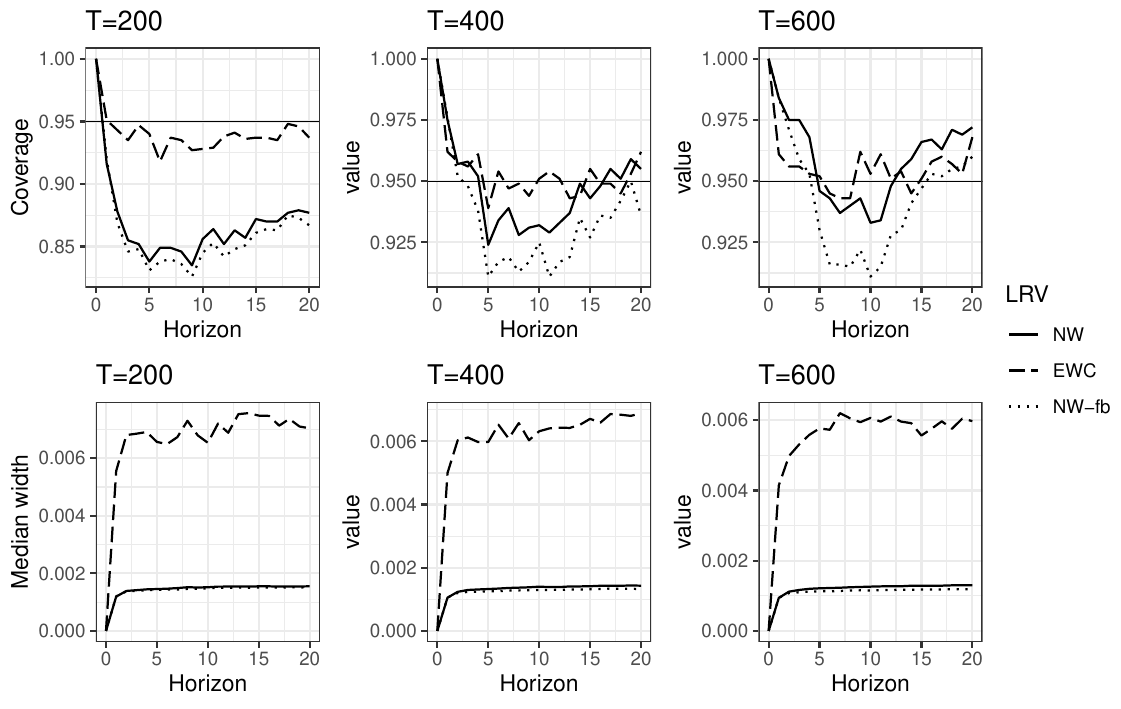}
\caption{Coverage rates and median interval widths of HDLP with different long-run covariance estimators in the sparse DFM.}\label{fig:sparse_lrvs}\label{fig:S6}
\end{figure}

We investigate the sensitivity of the proposed HDLP estimator to the choice of long-run covariance estimator. In addition to our default choice of Newey-West, we consider 
two alternative estimators suggested by \cite{lazarus2018har}, 
namely the
(i) Newey-West test ``NW-fb'' with recommended bandwidth choice $Q_T=1.3 T^{1/2}$ 
(Test 2 in Table 1 of \citealp{lazarus2018har}), and
(ii) Equal-weighted cosine test `EWC'' with  recommended number of cosine terms $\nu=0.4 T^{2/3}$ 
(Test 3 in Table 1 of \citealp{lazarus2018har}).
Both estimators are inconsistent, but can be used for valid inference using fixed-$b$ asymptotics. For the former, we use nonstandard critical values tabulated in Table 1 of \cite{kiefer_vogelsang_2005}. For the latter, we use critical values from the $t_{\nu}$ distribution. For further details on both estimators, see Section 2 of \cite{lazarus2018har} and references therein.
\Cref{fig:dense_lrvs} 
displays the coverage rates (top panels) and interval widths (bottom panels) for the proposed HDLP method with the three choices of long-run covariance estimator in the dense DFM simulation DGP, \Cref{fig:sparse_lrvs} for the sparse DFM.

In the dense DFM, we 
see 
that the  EWC estimator largely corrects for the poor coverage of the HDLP model with default NW estimator, giving coverage around 90\%  at horizon 1 and reaching nominal coverage around horizon 15.
The NW-fb estimator generally has slightly worse coverage than the default NW, except for the first horizon where both perform similarly.
Note that the poor performance of NW-fb compared to NW may appear contradictory with \cite{lazarus2018har}, e.g. Table 6 therein. This difference occurs likely due to the fact that we do not use  the ``textbook NW'' estimator of \cite{lazarus2018har} with bandwidth $Q_T=0.75T^{1/3}$ but instead  the adaptive bandwidth
$Q_T=1.1447(\hat\alpha(1)T)^{1/3}$, with $\hat\alpha(1)$ computed according to equation (6.4) in \cite{andrews1991heteroskedasticity}.
Looking at the median interval widths, we see that the improved coverage of the EWC comes at the cost of much wider intervals, being approximately 4-6 times as wide as the ones for the default NW across different sample sizes. The NW-fb intervals, in contrast, are slightly tighter, which likely contributes to the poorer coverage.

In the sparse DFM, the EWC estimator achieves the best coverage at nominal rates when $T=200$, compared to the regular NW at around 85\%. This difference narrows for the larger sample sizes $T=400$ and $600$ where both are slightly conservative in the 95-97\% coverage range. The NW-fb estimator performs generally worse than the regular NW. Besides, the relative interval sizes of the three estimators are  similar to the ones discussed in the dense DFM simulation design.

Overall, in the dense DFM where we expect the HDLP to suffer, the EWC results in near nominal coverage but comes at a high cost of wide intervals, unlike the NW.
In the sparse DFM, the coverage of HDLP with either NW or EWC is more similar-- especially at larger sample sizes --but the EWC's intervals still remaini much larger. Additionally, the performance of the NW-fb estimator is generally worse than our adaptive implementation of NW.

\subsection{Data Description}\label{sec:data}
\label{app:data}
\begin{table}[ht]
\centering
\caption{Definition of transformation codes}\label{fig:codes}
\begin{tabular}{lll}
\hline
    T& Transformation\\
\hline
    1& $f(x_t)=x_t$\\
    2& $f(x_t)=x_t-x_{t-1}$\\
    3& $f(x_t)=(x_t-x_{t-1})-(x_{t-1}-x_{t-2})$\\
    4& $f(x_t)=\log(x_t)$\\
    5& $f(x_t)=\log(x_t)-\log(x_{t-1})$\\
    6& $f(x_t)=(\log(x_t)-\log(x_{t-1}))-(\log(x_{t-1})-\log(x_{t-2}))$\\
   \hline
\end{tabular}
\end{table}

In the following tables, the data transformation codes follow the definitions in Table \ref{fig:codes}.
Codes marked with an asterisk indicate a transformation that is different from the default provided by \cite{McCrackenNg16} where we use the transformations in \cite{bernanke2005measuring} instead. The F/S column indicates whether we treat the variable as fast or slow respectively for our identification scheme.

\begin{table}[!ht]
\centering
\caption{Output \& Income  
}
\resizebox{0.85\textwidth}{!}{\makebox[\textwidth]{
\begin{tabular}{rlllcc}
  \hline
 & FRED & Description & DRI/McGraw & T & F/S \\ 
  \hline
1 & RPI & Real Personal Income & GMPYQ  & 5 & S \\ 
  2 & W875RX1 & Real personal income ex transfer receipts & GMYXPQ & 5 & S \\ 
  3 & INDPRO & IP Index & IP & 5 & S \\ 
  4 & IPFPNSS & IP: Final Products and Nonindustrial Supplies & IPP & 5 & S \\ 
  5 & IPFINAL & IP: Final Products (Market Group) & IPF & 5 & S \\ 
  6 & IPCONGD & IP: Consumer Goods & IPC & 5 & S \\ 
  7 & IPDCONGD & IP: Durable Consumer Goods & IPCD & 5 & S \\ 
  8 & IPNCONGD & IP: Nondurable Consumer Goods & IPCN & 5 & S \\ 
  9 & IPBUSEQ & IP: Business Equipment & IPE & 5 & S \\ 
  10 & IPMAT & IP: Materials & IPM & 5 & S \\ 
  11 & IPDMAT & IP: Durable Materials & IPMD & 5 & S \\ 
  12 & IPNMAT & IP: Nondurable Materials & IPMND & 5 & S \\ 
  13 & IPMANSICS & IP: Manufacturing (SIC) & IPMFG & 5 & S \\ 
  14 & IPB51222S & IP: Residential Utilities & IPUT & 5 & S \\ 
  15 & IPFUELS & IP: Fuels & - & 5 & S \\ 
  17 & CUMFNS & Capacity Utilization: Manufacturing & IPXMCA & 1* & S \\ 
   \hline
\end{tabular}
}}
\end{table}

\begin{table}[!ht]
\centering
\caption{Money \& Credit}
\resizebox{0.85\textwidth}{!}{\makebox[\textwidth]{
\begin{tabular}{rlllcc}
  \hline
 & FRED & Description & DRI/McGraw & T & F/S \\ 
  \hline
1 & M1SL & M1 Money Stock & FM1 & 5* & F \\ 
  2 & M2SL & M2 Money Stock & FM2 & 5* & F \\ 
  3 & M2REAL & Real M2 Money Stock & FM2DQ & 5 & F \\ 
  4 & BOGMBASE & St. Louis Adjusted Monetary Base & FMFBA & 5* & F \\ 
  5 & TOTRESNS & Total Reserves of Depository Institutions & FMRRA & 5* & F \\ 
  6 & NONBORRES & Reserves Of Depository Institutions & FMRNBA & 5* & F \\ 
  7 & BUSLOANS & Commercial and Industrial Loans & FCLNQ & 5* & F \\ 
  8 & REALLN & Real Estate Loans at All Commercial Banks & - & 6 & F \\ 
  9 & NONREVSL & Total Nonrevolving Credit & CCINRV & 5* & F \\ 
  10 & CONSPI & Nonrevolving consumer credit to Personal Income & - & 2 & F \\ 
  12 & DTCOLNVHFNM & Consumer Motor Vehicle Loans Outstanding & - & 6 & F \\ 
  13 & DTCTHFNM & Total Consumer Loans and Leases Outstanding & - & 6 & F \\ 
  14 & INVEST & Securities in Bank Credit at All Commercial Banks & - & 6 & F \\ 
   \hline
\end{tabular}
}
}
\end{table}

\begin{table}[!ht]
\centering
\caption{Labour Market}
\resizebox{0.85\textwidth}{!}{\makebox[\textwidth]{
\begin{tabular}{rlllcc}
  \hline
 & FRED & Description & DRI/McGraw & T &  F/S \\ 
  \hline
1 & HWI & Help-Wanted Index for United States & LHEL & 5* & S \\ 
  2 & HWIURATIO & Ratio of Help Wanted/No. Unemployed & LHELX & 4* & S \\ 
  3 & CLF16OV & Civilian Labor Force & LHEM & 5 & S \\ 
  4 & CE16OV & Civilian Employment & LHNAG & 5 & S \\ 
  5 & UNRATE & Civilian Unemployment Rate & LHUR & 1* & S \\ 
  6 & UEMPMEAN & Average Duration of Unemployment (Weeks) & LHU680 & 1* & S \\ 
  7 & UEMPLT5 & Civilians Unemployed - Less Than 5 Weeks & LHU5 & 1* & S \\ 
  8 & UEMP5TO14 & Civilians Unemployed for 5-14 Weeks & LHU14 & 1* & S \\ 
  9 & UEMP15OV & Civilians Unemployed - 15 Weeks \& Over & LHU15 & 1* & S \\ 
  10 & UEMP15T26 & Civilians Unemployed for 15-26 Weeks & LHU26 & 1* & S \\ 
  11 & UEMP27OV & Civilians Unemployed for 27 Weeks and Over & - & 5 & S \\ 
  12 & CLAIMSx & Initial Claims & - & 5 & S \\ 
  13 & PAYEMS & All Employees: Total nonfarm & LPNAG & 5 & S \\ 
  14 & USGOOD & All Employees: Goods-Producing Industries & LPGD & 5 & S \\ 
  15 & CES1021000001 & All Employees: Mining and Logging: Mining & LPMI & 5 & S \\ 
  16 & USCONS & All Employees: Construction & LPCC & 5 & S \\ 
  17 & MANEMP & All Employees: Manufacturing & LPEM & 5 & S \\ 
  18 & DMANEMP & All Employees: Durable goods & LPED & 5 & S \\ 
  19 & NDMANEMP & All Employees: Nondurable goods & LPEN & 5 & S \\ 
  20 & SRVPRD & All Employees: Service-Providing Industries & LPSP & 5 & S \\ 
  21 & USTPU & All Employees: Trade, Transportation \& Utilities & LPTU & 5 & S \\ 
  22 & USWTRADE & All Employees: Wholesale Trade & LPT & 5 & S \\ 
  23 & USTRADE & All Employees: Retail Trade & - & 5 & S \\ 
  24 & USFIRE & All Employees: Financial Activities & LPFR & 5 & S \\ 
  25 & USGOVT & All Employees: Government & LPGOV & 5 & S \\ 
  26 & CES0600000007 & Avg Weekly Hours : Goods-Producing & - & 1 & S \\ 
  27 & AWOTMAN & Avg Weekly Overtime Hours : Manufacturing & LPMOSA & 1* & S \\ 
  28 & AWHMAN & Avg Weekly Hours : Manufacturing & LPHRM & 1 & S \\ 
  30 & CES0600000008 & Avg Hourly Earnings : Goods-Producing & - & 6 & S \\ 
  31 & CES2000000008 & Avg Hourly Earnings : Construction & LEHCC & 5* & S \\ 
  32 & CES3000000008 & Avg Hourly Earnings : Manufacturing & LEHM & 5* & S \\ 
   \hline
\end{tabular}
}}
\end{table}

\begin{table}[!ht]
\centering
\caption{Consumption \& Orders}
\resizebox{0.85\textwidth}{!}{\makebox[\textwidth]{
\begin{tabular}{rlllcc}
  \hline
 & FRED & Description & DRI/McGraw & T &  F/S \\ 
  \hline
1 & HOUST & Housing Starts: Total New Privately Owned & HSFR & 4 & F \\ 
  2 & HOUSTNE & Housing Starts, Northeast & HSNE & 4 & F \\ 
  3 & HOUSTMW & Housing Starts, Midwest & HSMW & 4 & F \\ 
  4 & HOUSTS & Housing Starts, South & HSSOU & 4 & F \\ 
  5 & HOUSTW & Housing Starts, West & HSWST & 4 & F \\ 
  6 & PERMIT & New Private Housing Permits (SAAR) & - & 4 & F \\ 
  7 & PERMITNE & New Private Housing Permits, Northeast (SAAR) & - & 4 & F \\ 
  8 & PERMITMW & New Private Housing Permits, Midwest (SAAR) & - & 4 & F \\ 
  9 & PERMITS & New Private Housing Permits, South (SAAR)  & - & 4 & F \\ 
  10 & PERMITW & New Private Housing Permits, West (SAAR) & - & 4 & F \\ 
   \hline
\end{tabular}
}}
\end{table}

\begin{table}[!ht]
\centering
\caption{Orders \& Inventories}
\resizebox{0.85\textwidth}{!}{\makebox[\textwidth]{
\begin{tabular}{rlllcc}
  \hline
 & FRED & Description & DRI/McGraw & T &  F/S \\ 
  \hline
1 & DPCERA3M086SBEA & Real personal consumption expenditures & GMCQ & 5 & S \\ 
  2 & CMRMTSPLx & Real Manu. and Trade Industries Sales & - & 5 & F \\ 
  3 & RETAILx & Retail and Food Services Sales & - & 5 & F \\ 
  9 & AMDMNOx & New Orders for Durable Goods & - & 5 & F \\ 
  11 & AMDMUOx & Unfilled Orders for Durable Goods & - & 5 & F \\ 
  12 & BUSINVx & Total Business Inventories & - & 5 & F \\ 
  13 & ISRATIOx & Total Business: Inventories to Sales Ratio & - & 2 & F \\ 
   \hline
\end{tabular}
}}
\end{table}

\begin{table}[!ht]
\centering
\caption{Interest rate \& Exchange rates}
\resizebox{0.85\textwidth}{!}{\makebox[\textwidth]{
\begin{tabular}{rlllcc}
  \hline
 & FRED & Description & DRI/McGraw & T & F/S \\ 
  \hline
1 & FEDFUNDS & Effective Federal Funds Rate & FYFF & 1* & F \\ 
  2 & CP3Mx & 3-Month AA Financial Commercial Paper Rate & - & 2 & F \\ 
  3 & TB3MS & 3-Month Treasury Bill & FYGM3 & 1* & F \\ 
  4 & TB6MS & 6-Month Treasury Bill & FYGM6 & 1* & F \\ 
  5 & GS1 & 1-Year Treasury Rate & FYGT1 & 1* & F \\ 
  6 & GS5 & 5-Year Treasury Rate & FYGT5 & 1* & F \\ 
  7 & GS10 & 10-Year Treasury Rate & FYGT10 & 1* & F \\ 
  8 & AAA & Moody’s Seasoned Aaa Corporate Bond Yield & FYAAAC & 1* & F \\ 
  9 & BAA & Moody’s Seasoned Baa Corporate Bond Yield & FYBAAC & 1* & F \\ 
  10 & COMPAPFFx & 3-Month Commercial Paper Minus FEDFUNDS & - & 1 & F \\ 
  11 & TB3SMFFM & 3-Month Treasury C Minus FEDFUNDS & SFYGM3 & 1 & F \\ 
  12 & TB6SMFFM & 6-Month Treasury C Minus FEDFUNDS & SFYGM6 & 1 & F \\ 
  13 & T1YFFM & 1-Year Treasury C Minus FEDFUNDS & SFYGT1 & 1 & F \\ 
  14 & T5YFFM & 5-Year Treasury C Minus FEDFUNDS & SFYGT5 & 1 & F \\ 
  15 & T10YFFM & 10-Year Treasury C Minus FEDFUNDS & SFYGT10 & 1 & F \\ 
  16 & AAAFFM & Moody’s Aaa Corporate Bond Minus FEDFUNDS & SFYAAAC & 1 & F \\ 
  17 & BAAFFM & Moody’s Baa Corporate Bond Minus FEDFUNDS & SFYBAAC & 1 & F \\ 
  19 & EXSZUSx & Switzerland / U.S. Foreign Exchange Rate & EXRSW & 5 & F \\ 
  20 & EXJPUSx & Japan / U.S. Foreign Exchange Rate & EXRJAN & 5 & F \\ 
  21 & EXUSUKx & U.S. / U.K. Foreign Exchange Rate & EXRUK & 5 & F \\ 
  22 & EXCAUSx & Canada / U.S. Foreign Exchange Rate & EXRCAN & 5 & F \\ 
   \hline
\end{tabular}
}}
\end{table}

\begin{table}[!ht]
\centering
\caption{Prices}
\resizebox{0.85\textwidth}{!}{\makebox[\textwidth]{
\begin{tabular}{rlllcc}
  \hline
 & FRED & Description & DRI/McGraw & T & F/S \\ 
  \hline
1 & WPSFD49207 & PPI: Finished Goods & PWFSA & 5* & S \\ 
  2 & WPSFD49502 & PPI: Finished Consumer Goods & PWFCSA & 5* & S \\ 
  3 & WPSID61 & PPI: Intermediate Materials & PWIMSA & 5* & S \\ 
  4 & WPSID62 & PPI: Crude Materials & PWCMSA & 5* & S \\ 
  5 & OILPRICEx & Crude Oil, spliced WTI and Cushing & - & 6 & F \\ 
  6 & PPICMM & PPI: Metals and metal products & - & 6 & S \\ 
  8 & CPIAUCSL & CPI : All Items & PUNEW & 5* & S \\ 
  9 & CPIAPPSL & CPI : Apparel & PU83 & 5* & S \\ 
  10 & CPITRNSL & CPI : Transportation & PU84 & 5* & S \\ 
  11 & CPIMEDSL & CPI : Medical Care & PU85 & 5* & S \\ 
  12 & CUSR0000SAC & CPI : Commodities & PUC & 5* & S \\ 
  13 & CUSR0000SAD & CPI : Durables & PUCD & 5* & S \\ 
  14 & CUSR0000SAS & CPI : Services & PUS & 5* & S \\ 
  15 & CPIULFSL & CPI : All Items Less Food & PUXF & 5* & S \\ 
  16 & CUSR0000SA0L2 & CPI : All items less shelter & PUXHS & 5* & S \\ 
  17 & CUSR0000SA0L5 & CPI : All items less medical care & PUXM & 5* & S \\ 
  18 & PCEPI & Personal Cons. Expend.: Chain Index & - & 6 & S \\ 
  19 & DDURRG3M086SBEA & Personal Cons. Exp: Durable goods & GMCDQ & 5* & S \\ 
  20 & DNDGRG3M086SBEA & Personal Cons. Exp: Nondurable goods & GMCNQ & 5* & S \\ 
  21 & DSERRG3M086SBEA & Personal Cons. Exp: Services & GMCSQ & 5* & S \\ 
   \hline
\end{tabular}
}}
\end{table}

\begin{table}[!ht]
\centering
\caption{Stock Market}
\resizebox{0.85\textwidth}{!}{\makebox[\textwidth]{
\begin{tabular}{rlllcc}
  \hline
 & FRED & Description & DRI/McGraw & T & F/S\\ 
  \hline
1 & S\&P 500 & S\&P’s Common Stock Price Index: Composite & FSPCOM & 5 & F \\ 
  2 & S\&P: indust & S\&P’s Common Stock Price Index: Industrials & FSPIN & 5 & F \\ 
  3 & S\&P div yield & S\&P’s Composite Common Stock: Dividend Yield & FSDXP & 1* & F \\ 
  4 & S\&P PE ratio & S\&P’s Composite Common Stock: Price-Earnings Ratio & FSPXE & 1* & F \\ 
   \hline
\end{tabular}
}}
\end{table}

\clearpage
\subsection{FAVAR Implementation}\label{app:FAVAR}
We closely follow the method described in \cite{bernanke2005measuring} to estimate the FAVAR. 
The most important difference concerns the scaling of 
all impulse responses by the response of the FFR at horizon 0. That means the shock to which the variables are responding is of such a size
that the FFR rises by one on impact, as opposed to a size of one standard deviation, as is done by \cite{bernanke2005measuring}. We implement this change to 
ensures that the scale of the responses from the FAVAR are comparable to those from our HDLP specification. 

Let $\underset{720\times 122}{\bX_{\text{all}}}$ denote the matrix containing all variables, and $\underset{720\times 67}{\bX_{\text{slow}}}$ the one containing only the ``slow" variables which we assume to not react within the same period to a monetary policy shock. We then estimate the factors $\underset{720\times 3}{\hat \bC}$ and $\underset{720\times 3}{\hat\bC^*}$ as the first 3 principal components of $\bX_{\text{all}}$ and $\bX_{\text{slow}}$ respectively. We regress $\hat\bC$ onto $\hat\bC^*$ and $\underset{720\times 1}{\bR_s}$, where $\bR_s$ is the FFR ($\bR$) scaled such that it has mean 0 and variance 1, and estimate by OLS
\begin{equation*}
    \hat\bC=\left[\hat\bC^*, \bR_s\right]\underset{4\times 3}{\hat\bbeta}+\hat\bu.
\end{equation*}
We then take $\underset{1\times 3}{\hat\bb_R}$ as the last row of $\hat\bbeta$, and let $\underset{720\times 3}{\hat\bF}=\hat\bC-\bR_s\bb_R$. Next, we use the \texttt{vars} package in \texttt{R} to estimate a 13 lag VAR with the four variables in $[\hat\bF,\bR]$ (ordering the FFR last), and obtain the impulse responses  of the factors and $\bR$ using a recursive identification scheme with Cholesky decomposition, and store them in a matrix $\underset{(h_{\max}+1)\times 4}{\boldsymbol{IR_{\text{fac}}}}$. 
These impulse responses are all scaled such that we use the unit shock identification, similar to that of local projections. To obtain the impulse responses of the individual variables in $\bX_{\text{all}}$, we regress $\bX_{\text{all}}$ onto $\hat\bF$, and estimate by OLS
\begin{equation*}
    \bX_{\text{all}}=\hat\bF\underset{3\times122}{\hat\bLambda}+\hat\bv.
\end{equation*}
The matrix of impulse responses for all variables is then obtained by 
\begin{equation*}
    \underset{(h_{\max}+1)\times 122}{\boldsymbol{IR}}=\boldsymbol{IR_{\text{fac}}}\left[\begin{array}{c}
         \hat\bLambda\\
          \bzero
    \end{array}\right].
\end{equation*}
Finally, we cumulate the impulse responses for the variables which were taken in differences to obtain the response of the level variable. 

For inference, we use a simple residual bootstrap. We draw with replacement from the VAR residuals, and use them to construct the bootstrap series $\hat\bF^*$ and $\bR^*$. 
We then perform the same steps, i.e.\ estimate the VAR on the bootstrap data, obtain (unit-shock scaled) impulse responses for the factors,
compute the (cumulated) implied impulse responses of all variables.
We repeat this process $B=499$ times and use the 2.5\% and 97.5\% quantiles to construct confidence intervals.

\end{document}